\documentstyle[12pt,psfig]{article}

\begin{document}
\vskip 1.5cm
\centerline{\large\bf Non-equilibrium initial conditions}
\centerline{\large\bf from pQCD for RHIC and LHC}
\medskip
\bigskip
\medskip
\centerline{\bf N. Hammon, H. St\"ocker, W. Greiner \footnote{This work was supported
by BMBF, DFG, and GSI}}
\medskip
\bigskip
\centerline{Institut F\"ur Theoretische Physik}
\centerline{Robert-Mayer Str.~10}
\centerline{Johann Wolfgang Goethe-Universit\"at}
\centerline{60054 Frankfurt am Main}
\centerline{Germany}
\medskip
\bigskip
\medskip
\bigskip
\bigskip
\centerline{Abstract}
\bigskip
{\small We calculate the initial non-equilibrium conditions from 
perturbative QCD (pQCD) within Glauber multiple scattering theory for 
$\sqrt s =200$ AGeV and $\sqrt s =5.5$ ATeV. 
At the soon available collider energies
one will particularly test the small $x$ region of the parton distributions
entering the cross sections. Therefore shadowing effects, previously more or less 
unimportant,
will lead to new effects on variables such as particle multiplicities $dN/dy$, 
transverse energy 
production $d\bar{E}_T/dy$, and the initial temperature $T_i$.
In this paper we will have a closer look on the effects of shadowing 
by employing different parametrizations for the shadowing effect for valence quarks,
sea quarks and gluons.
Since the cross sections at midrapidity are dominated by processes involving gluons
the amount of their depletion is particularly important. We will therefore have a closer 
look on the results for $dN/dy$, $d\bar{E}_T/dy$, and $T_i$ by using two different
gluon shadowing ratios, differing strongly in size. As a matter 
of fact, the calculated quantities differ significantly.}
\newpage

\centerline{\bf 1. Introduction}

One of the challenging goals of heavy ion physics is the detection of the quark-gluon
plasma, a state in which the partons are able to move freely within a distance
larger than the typical confinement scale $r_{conf.}\sim 1/\Lambda_{QCD}\sim 1/0.2$ GeV 
$\sim 1$ fm. The build-up of this state should happen early in a heavy ion reaction
when the two streams of initially cold nuclear matter pass through each other.
Thereby first virtual partons are transformed to real ones and later on in the expansion 
phase the fragmentation of the partons into colorless hadrons takes place. When
separating pQCD from non-perturbative effects at some semi-hard scale $p_0 = 2$ GeV the
respective time scale of perturbative processes is thus of the order $\tau \sim 1/p_0 \sim
0.1$ fm/c which approximately coincides with the lower bound of the initial formation 
time of the 
plasma in a local cell \cite{hwa1}. Therefore all further evolution of the system is 
significantly influenced by the initial conditions of pQCD since macroscopic parameters, 
as e.g.~the initial temperature $T_i$, directly enter into hydrodynamical calculations.\\
We here will focus on the very early phase of an ultrarelativistic heavy ion collision
and use pQCD above the semi-hard scale $p_{sh.}=p_0 = 2$ GeV.\\

In a typical high energy $pp$ or $p\bar p$ event one measures distinct hadronic jets with 
a transverse momenta of several GeV ($p_T\geq 5$ GeV) \cite{ua1}. In contrast to the 
experimental very clean 
situation of hadronis jets at large $p_T$ one encounters the problem of detectability of low
transverse momentum jets in heavy ion collisions. These so-called minijets contribute
significantly to the transverse energy produced in AB collisions due to their large
multiplicity \cite{eskola1}. The major part of these set-free partons are gluons that
strongly dominate the processes as their number is much larger for the relevant momentum 
fractions. In turn the shadowing effects are expected to be much larger for gluons than
for the quark sea \cite{nils1}. Therefore the relative contribution of the gluons should 
decrease but still dominate the cross sections. The shadowing of the gluons has the
peculiarity of not being known exactly due to the neutrality of the mediators of the strong
interaction which makes it impossible to access $R_G(x,Q^2)$ directly in a deep inelastic 
$e+A$ event. Therefore we will here investigate two possible parametrizations of the 
shadowing ratio $R_G=xG^A/A\cdot xG^N$ for gluons as will be described below in detail.\\
\centerline{\bf 2. Minijets}
As outlined above, we will here investigate the effects of shadowing on the minijet
production cross sections. The production of a parton $f=g,q,\bar q$ can in leading 
order be described as \cite{eskola1}
\begin{eqnarray}
\frac{d\sigma ^f}{dy} &=& \int dp_{T}^{2}~ dy_2~ \sum_{ij,kl} ~x_1 f_{i}(x_1,Q^2)~ x_2 
f_{j}(x_2,Q^2) \nonumber \\
&\times & \left [ \delta_{fk} \frac{d\hat \sigma^{ij\rightarrow kl}}{d\hat t} (\hat t, \hat u)+
\delta_{fl} \frac{d\hat \sigma^{ij\rightarrow kl}}{d\hat t}(\hat u, \hat t) \right ]
\frac{1}{1+\delta_{kl}}
\end{eqnarray}
The factor $1/(1+\delta_{kl})$ enters due to the symmetry of processes with 
two identical partons in the final state. The exchange term 
$d\hat \sigma (\hat t, \hat u) \leftrightarrow d\hat \sigma (\hat u, \hat t)$ accounts for the 
possible symmetries of e.g.~having a quark from nucleon $i$ and a gluon from nucleon $j$
and vice versa, i.e.~it handles the interchange of two of the propagators in the
scattering process.
The possible combinations of initial states are
\begin{equation}
ij = gg,gq,qg,g\bar q,\bar q g, qq, q\bar q, \bar q q, \bar q\bar q
\end{equation}
The momentum fractions of the partons in the initial state are 
\begin{equation}
x_1 = \frac{p_T}{\sqrt s} \left [e^y +e^{y_2}\right ],~~~
x_2 = \frac{p_T}{\sqrt s} \left [e^{-y} +e^{-y_2}\right ]
\end{equation}
The integration regions are
\begin{equation}
p_{0}^{2} \leq p_{T}^{2} \leq \left ( \frac {\sqrt s}{2 {\rm cosh} ~y} \right)^2,~~~
-{\rm ln} \left( \frac{\sqrt s}{p_T}-e^{-y}\right) \leq y_2 \leq 
{\rm ln} \left( \frac{\sqrt s}{p_T}-e^{-y}\right)
\end{equation}
with 
\begin{equation}
\left | y \right | \leq {\rm ln} \left ( \frac {\sqrt s}{2 p_0} + 
\sqrt{ \frac{s}{4 p_{0}^{2}}-1} \right )
\end{equation}
The mandelstam variables are defined as
\begin{equation}
\hat s =x_1\cdot x_2 \cdot s, ~\hat t = -p_{T}^2 \left[ 1+e^{(y_2 -y)}\right ],~
\hat u = -p_{T}^2 \left[ 1+e^{(y -y_2)}\right ]
\end{equation}
For the parton distributions entering the handbag graph we choose the GRV LO set 
\cite{grv} for RHIC. Since at LHC one probes smaller momentum fractions we
there use the newer CTEQ4L parametrization \cite{cteq} with $N_f=4$ and $Q=p_T$. 
The normalization is done so that one has two outgoing partons in one collision, i.e.
\begin{equation}
\int dy \frac{d\sigma ^f}{dy} = 2\sigma _{hard}^{f}
\end{equation}
In the calculations the boundaries for the calculations are either over the whole
rapidity range or $\left| y\right| \leq 0.5$ for the central rapidity region.\\
To account for the higher order contributions at RHIC we choose a fixed K factor of 
K=2.5 from comparison with experiment as discussed in \cite{eskola2,ua1}. 
In the range $5.5$ GeV 
$\leq p_T \leq 25$ GeV a factor K=2.5 is needed to describe the UA1 data, and in the range
$30$ GeV $\leq p_T \leq 50$ GeV a factor of K=1.6 is needed. However the cross section has 
dropped so much at these large transverse momenta that we keep K=2.5 fixed for all $p_T$.
For LHC energies the mean $p_T$ tends to be larger; so we choose K=1.5 for this case.\\
By applying Glauber theory we calculate the mean number of {\it events} per unit of 
rapidity:
\begin{equation}
\frac{dN^f}{dy} = T_{AA}(b) \frac{d\sigma^{f}_{hard}}{dy} 
\end{equation}
where the nuclear overlap function $T_{AA}(b)$ for central events is given by
$T_{AA}(0)\approx A^2/\pi R_{A}^2$. For the nuclei in our calculation this gives
$T_{AuAu}(0) = 29/mb$ and $T_{PbPb}(0) = 32/mb$. Again it should be emphasized that 
$dN^f/dy$ gives the number of collisions and that the total number of partons is as 
twice as 
high in a $2\rightarrow 2$ process. The necessary volume, needed to derive the densities 
from the absolute numbers, is calculated as
\begin{equation}
V_i = \pi R_{A}^2 \Delta y /p_0,~~R_{A}= A^{1/3} \times 1.1~fm
\end{equation}
Therefore we get $V_i (Au+Au)=12.9~fm^3$ and $V_i (Pb+Pb)=13.4~fm^3$.\\
For the energy density at midrapidity we need the first $E_T$ moment:
\begin{eqnarray}
\sigma^f \left < E_T\right > &=& \int dE_T \frac{d\sigma ^f}{dE_T} \left < E_T\right >\nonumber\\
&=& \int dp_{T}^{2}~ dy~ dy_2~ \sum_{ij,kl} ~x_1 f_{i}(x_1,Q^2)~ x_2 f_{j}(x_2,Q^2) \nonumber\\
&\times & \left [ \delta_{fk} \frac{d\hat \sigma^{ij\rightarrow kl}}{d\hat t} (\hat t, \hat u)+  
\delta_{fl} \frac{d\hat \sigma^{ij\rightarrow kl}}{d\hat t}(\hat u, \hat t) \right ]
\frac{1}{1+\delta_{kl}} ~ p_T ~\epsilon (y)
\end{eqnarray}
Here the acceptance function $\epsilon (y)$ is $\epsilon (y)=1$ for $\left| y\right|\leq 0.5$
and $\epsilon (y)=0$ otherwise.\\
\centerline{\bf 3. Nuclear Shadowing}
In heavy ion collisions one has to account for an effect that does not appear for
processes involving two nucleons only: nuclear shadowing.
In the lab frame the deep inelastic scattering at small Bjorken
$x$ ($x\ll 0.1$) proceeds via the vector mesons as described in the vector meson 
dominance model (VMD) 
where the handbag graph contribution becomes small. In VMD the interaction of the 
virtual photon with a nucleon or nucleus is described as a two step process: the photon
fluctuation into a $q\bar q$ pair (the $\rho,\omega,\phi$ mesons at small $Q^2$) 
within the coherence time $l_c$ and a subsequent strong 
interaction with the target \cite{bauer}. 
\begin{figure}
\centerline{\psfig{figure=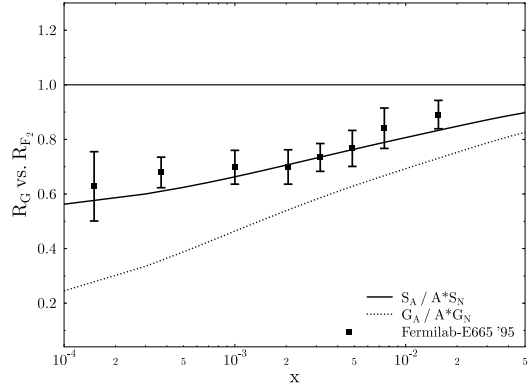,width=10cm}}
\vspace*{-1.5cm}
\caption{$R_{F_2}$ vs. $R_G$ at $Q^2 = 4$ GeV$^2$ for $^{207}Pb$.}
\label{ratio}
\end{figure}
The coherence time arises in this picture from the longitudinal momentum shift
between the photon and the fluctuation: $l_c \approx 1/\Delta k_z$
where $\Delta k_z = k_z^\gamma -k_z^h$. The cross section is:
\begin{equation}
\sigma (\gamma^{*}N)=\int_{0}^1 dz \int d^2 r \left|
\psi(z,r)\right|^2\sigma_{q\bar q N}(r)
\end{equation}
where the Sudakov variable $z$ gives the momentum fraction carried by the quark
or the antiquark. The 
interaction of the fluctuation with the nucleon can be described 
in the color transparency model as \cite{lonya1}
\begin{equation}
\sigma_{q\bar q N} = \frac{\pi^2}{3}r^{2}\alpha _s(Q'^2) x' g (x', Q'^2)
\end{equation}
where $x'=M_{q\bar q}^{2}/(2m\nu)$, $r$ is the transverse separation of
the pair and $Q'^2 =4/r^2$. For the interaction of the fluctuation with a nucleus 
one makes use of Glauber-Gribov multiple scattering theory \cite{gribov} where the 
fluctuation interacts coherently with more than one nucleon in the nucleus when
the coherence length exceeds the mean separation between two nucleons:
\begin{equation}
\sigma_{q\bar q A}=\int d^2b\left( 1-e^{-\sigma_{q\bar qN} T_A(b)/2} \right)
\end{equation}
\begin{figure}
\centerline{\psfig{figure=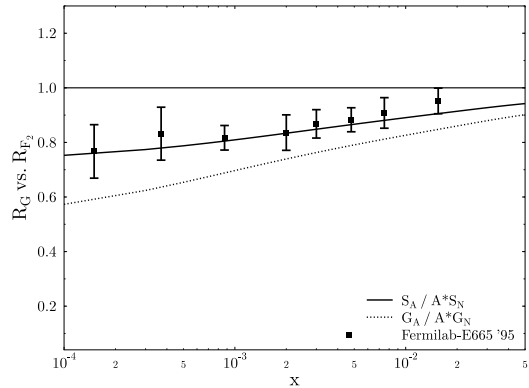,width=10cm}}
\vspace*{-1.5cm}
\caption{$R_{F_2}$ vs. $R_G$ at $Q^2 = 4$ GeV$^2$ for $^{40}Ca$.}
\label{ratio2}
\end{figure} 
When expanding for large nuclei and taking the dominating double scattering term only
one finds
\begin{equation}
\sigma_{hA} = A \sigma_{hN} \left[ 1- A^{1/3} \frac{\sigma_{hN}}{8\pi a^2}+\dots \right]
\end{equation}
with $a=1.1fm$.\\
Figures \ref{ratio} and \ref{ratio2} show the results for $^{207}Pb$ and $^{40}Ca$
(for further details see \cite{nils1}).\\
A very different scenario is employed in parton fusion models. Here the process
of parton parton fusion in nuclei can be understood as an overlapping of quarks 
and gluons that yields a reduction of
number densities at small $x$ and a creation of antishadowing for momentum conservation
at larger $x$ \cite{close}. The onset of this fusion process can be estimated to start
at values of the momentum fraction where the longitudinal wavelength ($1/xP$)
of a parton exceeds the size of a nucleon (or the inter-nucleon distance) inside
the Lorentz contracted nucleus: $1/xP \approx 2 R_n M_n /P$, corresponding to a value
$x \approx 0.1$. Originally the idea of parton fusion was proposed in \cite{glr} and 
later proven in \cite{mq} to appear when the total transverse size $1/Q$ of the partons in 
a nucleon becomes larger than the proton radius to yield a transverse overlapping within
a unit of rapidity, $xG(x)\geq Q^2 R^2$. The usual gluon distribution in the nucleon 
on the light cone in light-cone gauge ($n\cdot A =A^+ =0$) is given by
\begin{equation}
x G(x) = -(n^-)^2 \int \frac{d\lambda}{2\pi} \left< P\left| F^{+\mu} (0)
F^{+}\,_{\mu}(\lambda n)\right| P\right>
\end{equation}
The recombination is then described as the fusion of two gluon ladders into a single vertex.
One finally arives at a modified Altarelli-Parisi equation where the fusion correction
enters as a twist four light cone correlator. Typically the fusion correction in the free
nucleon turns
out to be significant only for unusually small values of $x$ or $Q^2$.
As shown in \cite{eskola5} the situation changes dramatically in heavy nuclei.
Here the strength of the fusion  for ladders coming from independent constituents increases
and is of the same order as the fusion from non-independent constituents. Therefore, parton
recombination is strongly increased in heavy nuclei of $A\sim 200$.\\

\begin{figure}
\vspace{-3cm}
\centerline{\psfig{figure=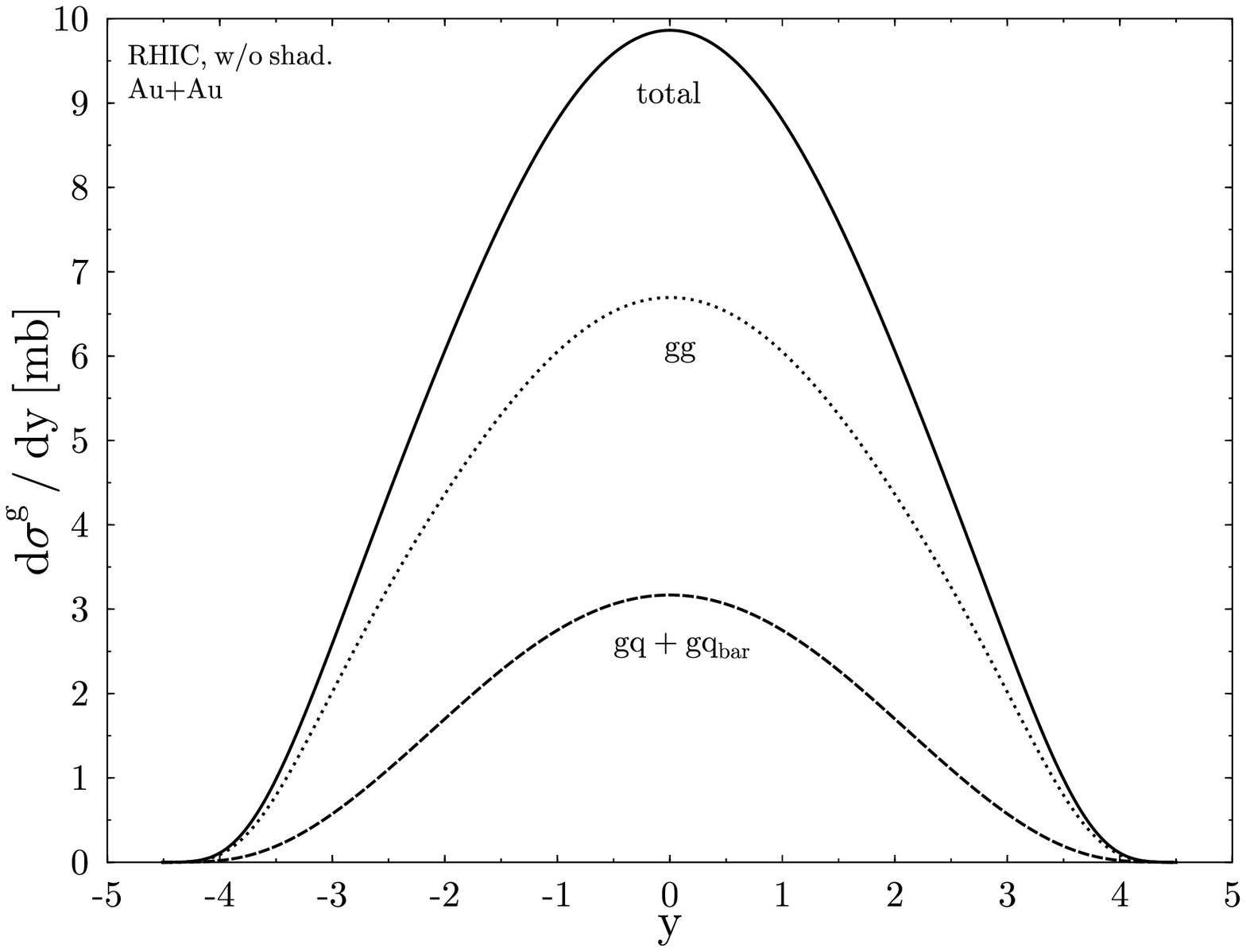,width=11cm}}
\vspace*{-1cm}
\centerline{\psfig{figure=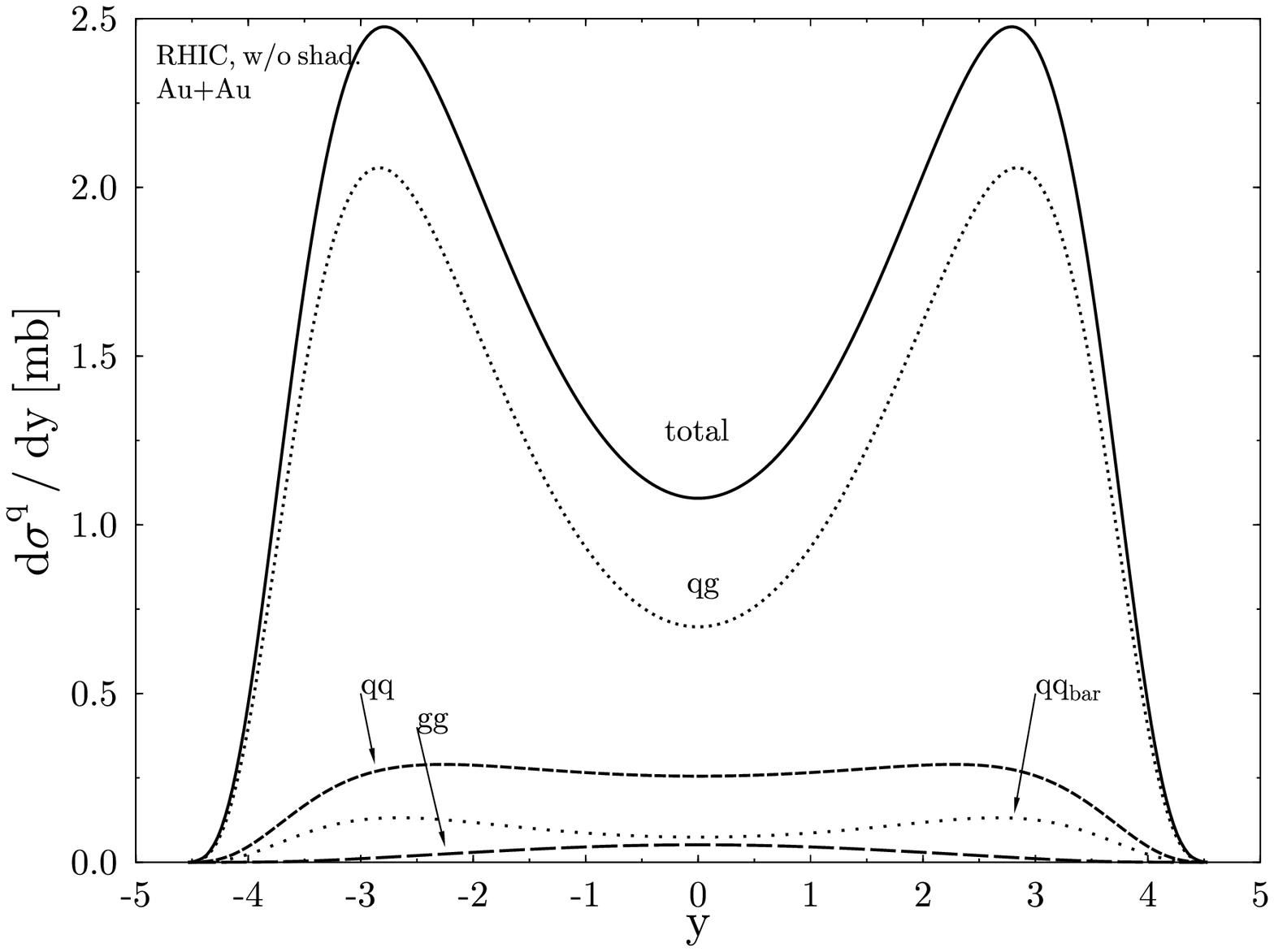,width=11cm}\hspace{-3cm}\psfig{figure=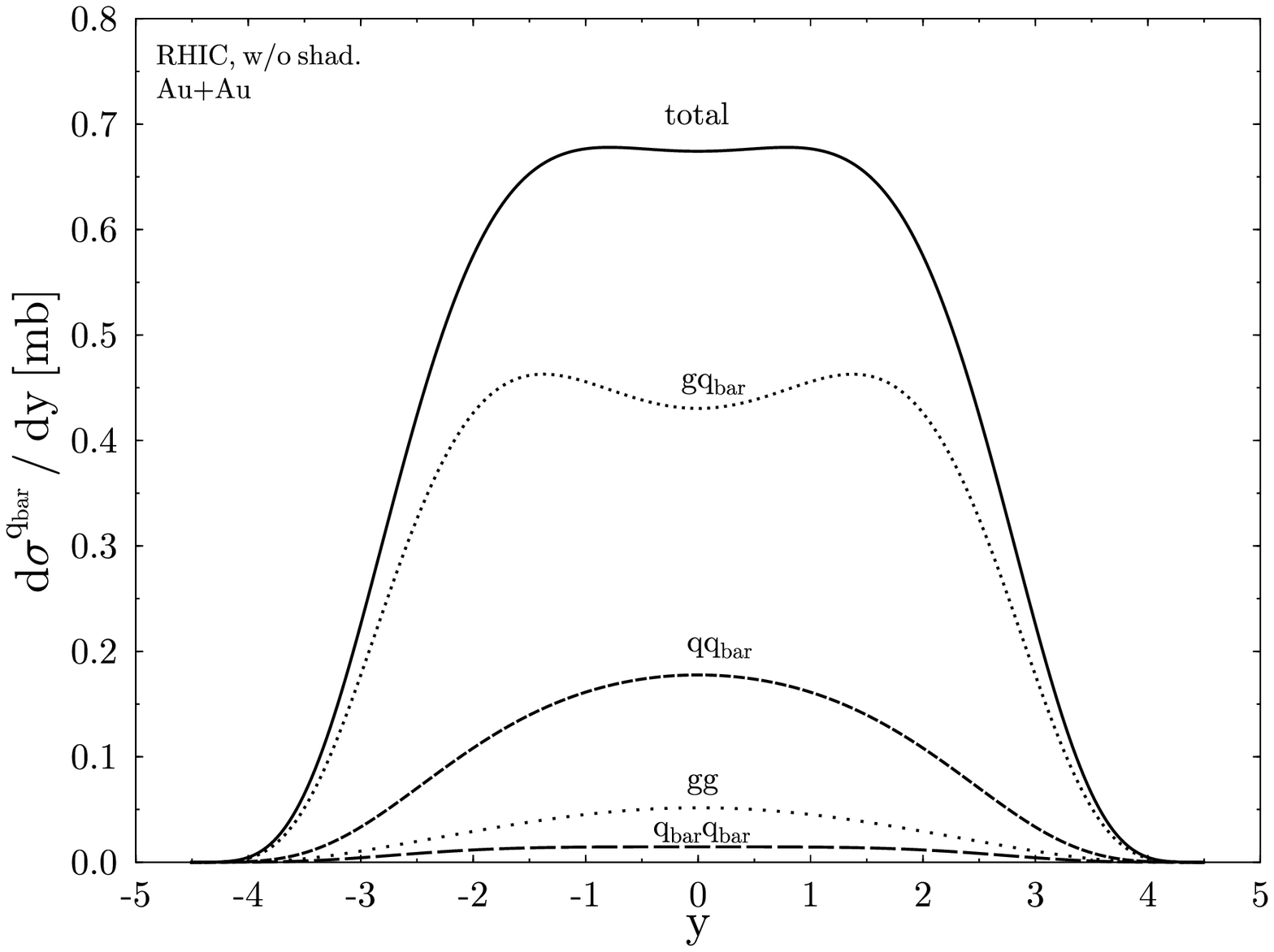,width=11cm}}
\vspace*{-1cm}
\caption{Unshadowed rapidity distributions of gluons, quarks, and antiquarks.}
\label{fig-rhic-noshad}
\end{figure}

\begin{figure}
\vspace{-2cm}
\centerline{\psfig{figure=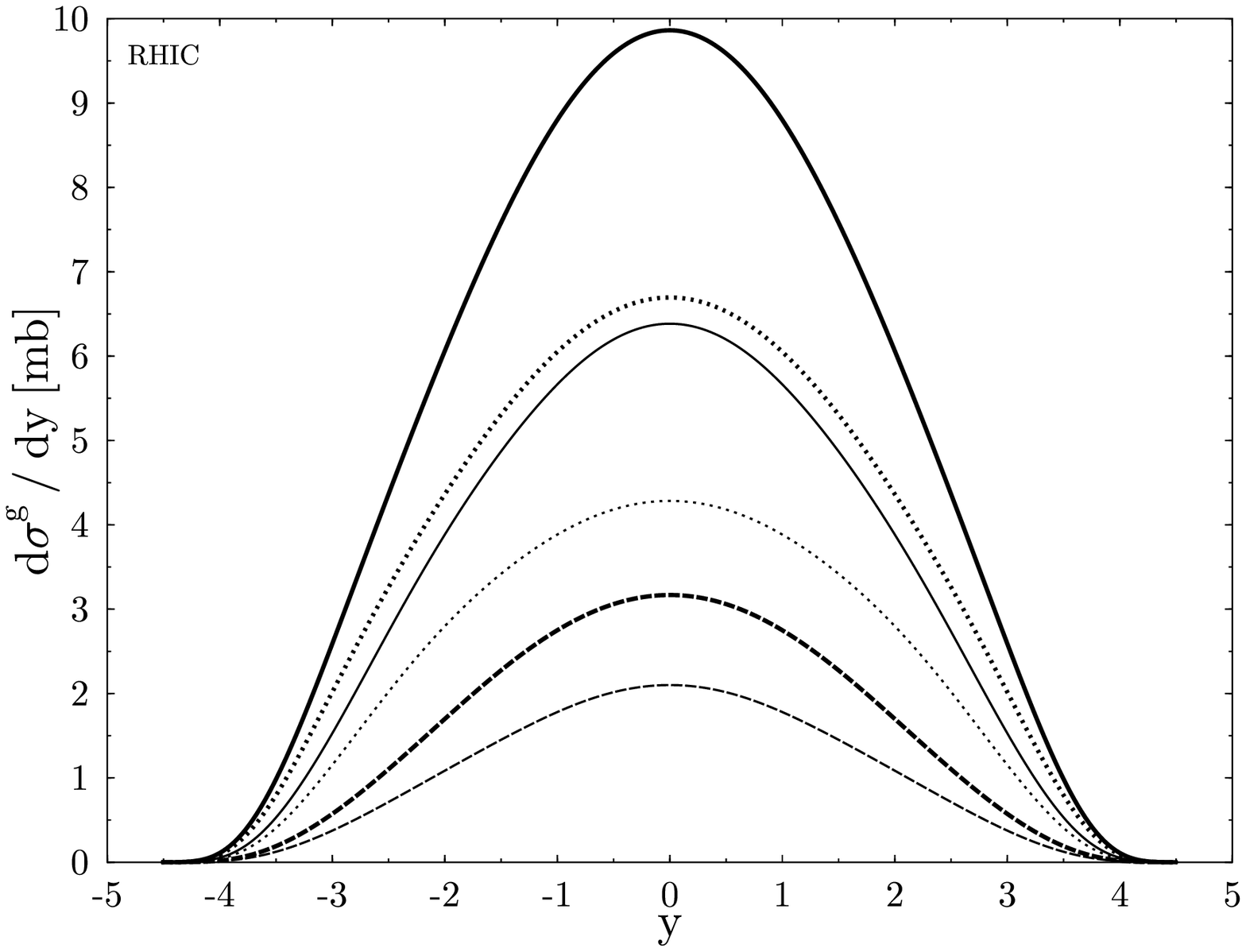,width=11cm}\hspace{-3.2cm}\psfig{figure=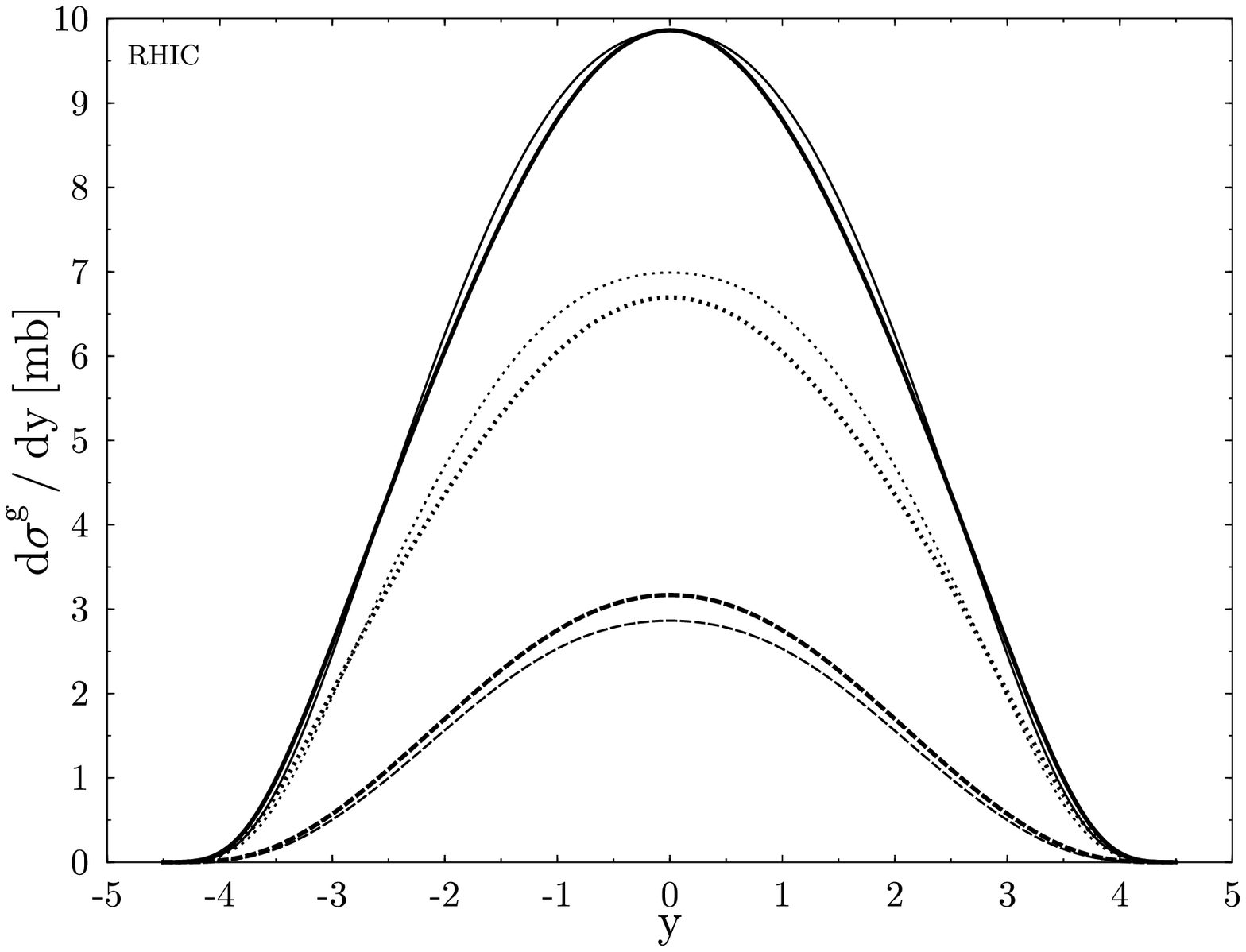,width=11cm}}
\vspace*{-1cm}
\caption{Comparison of the rapidity distributions with strong gluon shadowing (left figure)
and weak shadowing (right figure) to unshadowed distribution for RHIC (see text).}
\label{rhic-shadvsnoshad}
\end{figure}

\begin{figure}
\vspace{-2cm} 
\centerline{\psfig{figure=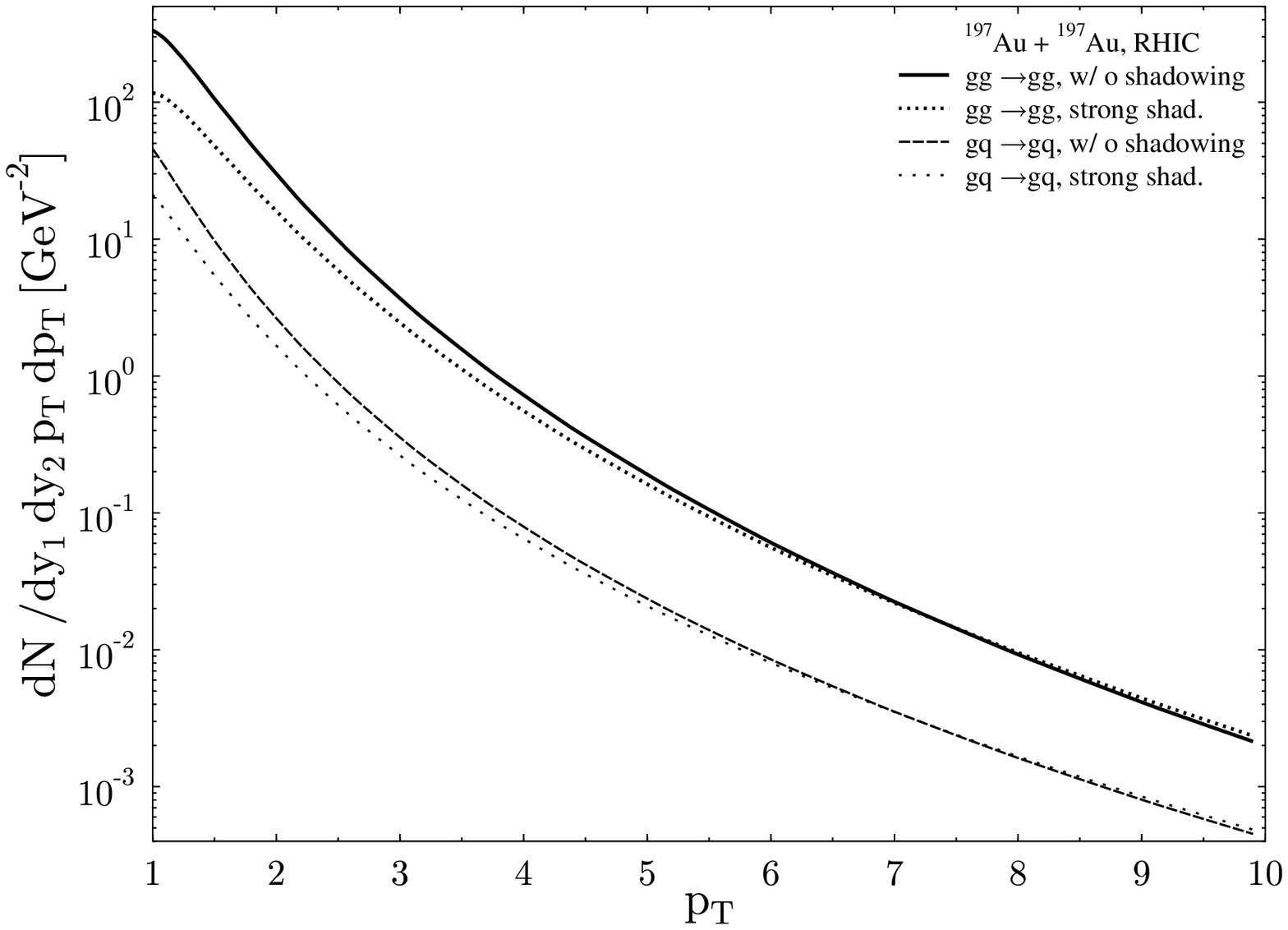,width=11cm}\hspace{-3.2cm}\psfig{figure=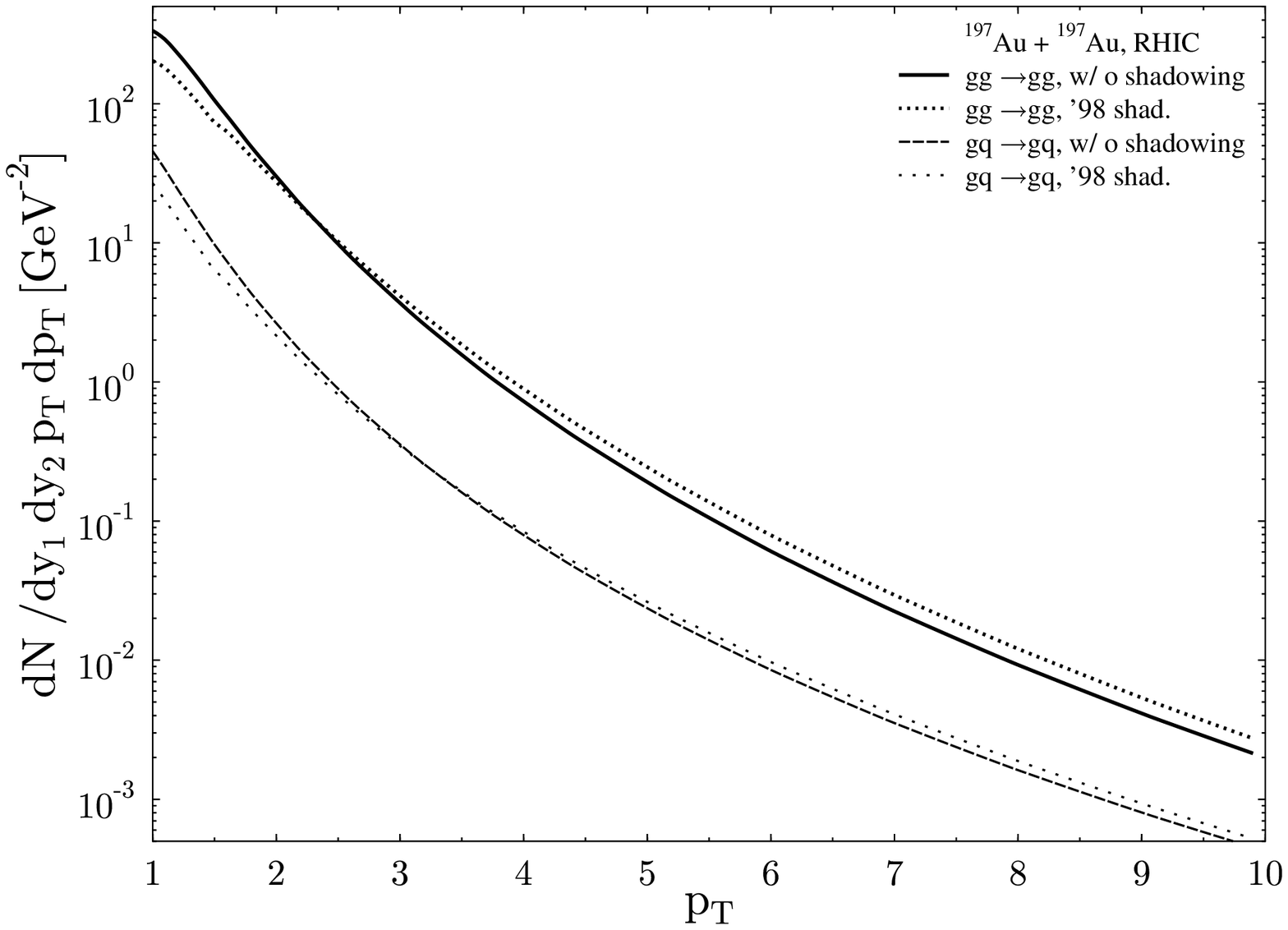,width=11cm}}
\vspace*{-1.5cm}
\caption{$p_T$ distributions for the two shadowing parametrizations at midrapidity.}
\label{rhic-ptdist}
\end{figure}

Unfortunately the different models do not give the same results for the ratio $R_G(x,Q^2)$.
We will therefore use two versions of parametrizations to investigate the effects 
of shadowing on the relevant variables. 
\begin{figure}
\centerline{\psfig{figure=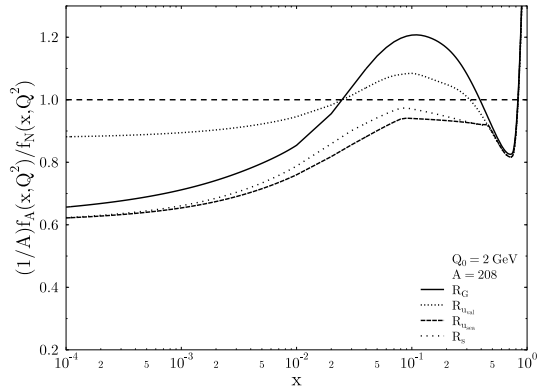,width=10cm}}
\vspace*{-1.5cm}
\caption{Shadowed parton distributions as parametrized by Eskola et al
in \protect\cite{eskola3}. Note that the
gluon shadowing appears to be weaker than quark shadowing and that the onset happens
for smaller $x$.}
\label{karisshad}
\end{figure}
On the one hand we use a $Q^2$ dependent version of Eskola, Kolhinen, Salgado, and 
Ruuskanen, often referred to as "'98 shadowing"
(see figure \ref{karisshad}), that tries to avoid any model dependence by using sum 
rules for baryon 
number and momentum \cite{eskola3} and on the other hand we use a modified version of 
a $Q^2$ independent parametrization (see figure \ref{nilsshad}) given in 
\cite{eskola4} which employs a much stronger gluon shadowing in accordance with the 
results of \cite{nils1}.
Especially for RHIC, where the lower bound for the momentum fraction at midrapidity 
for $p_T=p_0=2$ GeV is given by $x=2p_T/\sqrt s=0.02$, the onset of the gluon shadowing,
i.e.~the transition region between shadowing and antishadowing, is of great importance.\\
In \cite{eskola3} the onset of gluon shadowing ($R_G=1$) is chosen at 
$x\approx 0.029$ for $Q=2$ GeV motivated by the results found in \cite{gousett}
where the connection between the gluon distribution and the $Q^2$ dependence of $F_2$
via the DGLAP equations was employed:
\begin{figure}
\centerline{\psfig{figure=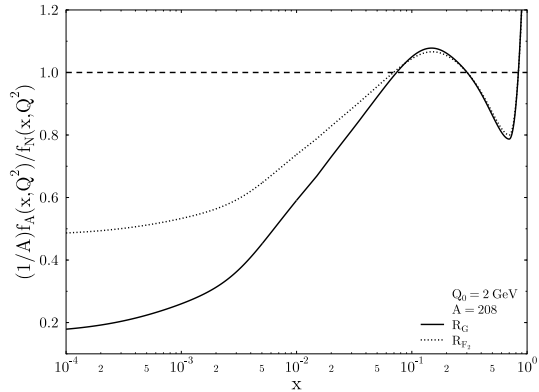,width=10cm}}
\vspace*{-1.5cm}
\caption{Our variation of the parametrization given in \protect\cite{eskola4} with much 
stronger gluon shadowing as found in the VMD calculation in \protect\cite{nils1}.
The stronger shadowing is also motivated by the fact that we calculate central collisions 
where the shadowing effect is stronger than for $b$-averaged collisions.}
\label{nilsshad}
\end{figure}
\begin{equation}
\frac{\partial F_2}{\partial {\rm ln}~Q^2}\sim \sum_i e_{i}^{2}xG(2x,Q^2)
\end{equation}
By using the NMC data \cite{nmc} on deep inelastic scattering on a combination of 
$Sn$ and $C$ targets the ratio $G^{Sn}(x)/G^{C}(x)$ was derived in the range
$0.011\leq x\leq 0.18$. The cross over point can, despite the large errorbars, be
guessed to be $x\approx 0.03$. However one should add here that the situation for 
$R_{G}^{Pb}=xG^{Pb}(x)/xG^{N}(x)$ can look rather different. 
Since this question of the onset of gluon 
shadowing is not yet settled we chose the same onset for quark and gluon shadowing
in our modified parametrization to investigate the relevance of this point. We fixed
$R_{F_2}=R_G=1$ at $x\approx 0.07$. In VMD as well as in parton fusion models the
onset is treated on an equal footing: for the coherent scattering processes in VMD 
it should make no difference (at least for the onset) whether a $q\bar q$ or a $gg$ 
pair scatters from more than one nucleon at $l_c\geq r_{NN}$. 
In the parton fusion model one treats the leaking out
of the partons equally for sea quarks and for gluons since for both sea quarks and for gluons
one has a spatial extent of $1/xP$ in the longitudinal direction and therefore the 
onset for $R_G$ and $R_{F_2}$ is essentially the same in this model.\\
\centerline{\bf 4. Results}
In the following we will give the results 
for the different parton species $f=g, q, \bar q$ at
RHIC and LHC including the different shadowing parametrizations or none shadowing,
respectively. The results for the number of partons $\int dN^f /dy$ can easily be derived
from $\int dy ~d\sigma ^f /dy$ by the relation $dN^f/dy =T_{AA}(0)d\sigma ^f/dy$.
All results include a K-factor of K=2.5 for RHIC and K=1.5 for LHC.
On the one hand we give the results for the whole $y$-range
and on the other hand we give the result for the central rapidity region which is of
special interest, not only from the experimental setup point of view but also since
it is the region where highest parton densities and the strongest shadowing effects 
are expected.\\
Let us start by giving the results without shadowing corrections for RHIC.
The first three tables give the unshadowed multiplicities integrated over the whole 
rapidity range and over the central region, respectively. Tables 4 through 6 give the 
first $E_T$ moments for the respective parton species. 
The rapidity distributions for the cross sections are depicted in 
figure \ref{fig-rhic-noshad}.\\ \\
{\bf Table 1}:$\int dy ~dN^g /dy$ {\it for} $\sqrt s =200$ {\it AGeV} \\
\begin{tabular}{|c||c|c|c|} \hline
range of $y$ & $gg\rightarrow gg$ & $gq\rightarrow gq$
+ $g\overline q\rightarrow g\overline q$ & TOTAL\\ \hline \hline
all $y$ & 920.8 & 384.3 & 1305.1 \\ \hline
$\left| y\right| \leq 0.5$ & 192.9 & 90.7 & 283.6 \\ \hline
\end{tabular}
\vspace*{1cm}\\
\noindent
{\bf Table 2}: $\int dy ~dN^q /dy$ {\it for} $\sqrt s =200$ {\it AGeV} \\
\begin{tabular}{|c||c|c|c|c|c|} \hline
range of $y$ & $gq\rightarrow gq$ & $qq\rightarrow qq$ &
$gg \rightarrow q\overline q$ & $q\overline q \rightarrow q\overline q$ &
TOTAL\\ \hline \hline
all $y$ & 310.3 & 57.3 & 6.5 & 22.8 & 396.9 \\ \hline
$\left| y\right| \leq 0.5$ & 21.0 & 7.4 & 1.5 & 2.3 & 32.2 \\ \hline
\end{tabular}
\vspace*{1cm}\\
%
{\bf Table 3}: $\int dy ~dN^{\bar q}/dy$ {\it for} $\sqrt s =200$ {\it AGeV} \\
\begin{tabular}{|c||c|c|c|c|c|} \hline
range of $y$ & $g\bar q\rightarrow g\bar q$ & $q\bar q\rightarrow q\bar q$ &
$gg \rightarrow q\overline q$ & $\bar q\bar q \rightarrow \bar q\bar q$ &   
TOTAL\\ \hline \hline
all $y$ & 74.2 & 22.8 & 6.5 & 2.2 & 105.7 \\ \hline
$\left| y\right| \leq 0.5$ & 12.5 & 5.3 & 1.5 & 0.4 & 19.7 \\ \hline
\end{tabular}
\vspace*{1cm}\\
\cleardoublepage
{\bf Table 4}:$\sigma^{g} \left< E_T\right>$ [mb GeV]\\
\begin{tabular}{|c||c|c|c|} \hline
range of $y$ & $gg\rightarrow gg$ & $gq\rightarrow gq$
+ $g\overline q\rightarrow g\overline q$ & TOTAL\\ \hline \hline
$\left| y\right| \leq 0.5$ & 18.02 & 8.72 & 26.74 \\ \hline
\end{tabular}
\vspace*{1cm}\\
{\bf Table 5}:$\sigma^{q} \left< E_T\right>$ [mb GeV]\\
\begin{tabular}{|c||c|c|c|c|c|} \hline
range of $y$ & $gq\rightarrow gq$ & $qq\rightarrow qq$ &
$gg \rightarrow q\overline q$ & $q\overline q \rightarrow q\overline q$ &
TOTAL\\ \hline \hline
$\left| y\right| \leq 0.5$ & 2.065 & 0.786 & 0.1398 & 0.22 & 3.2 \\ \hline
\end{tabular}
\vspace*{1cm}\\
{\bf Table 6}:$\sigma^{\bar q} \left< E_T\right>$ [mb GeV]\\
\begin{tabular}{|c||c|c|c|c|c|} \hline
range of $y$ & $g\bar q\rightarrow g\bar q$ & $\bar q\bar q\rightarrow 
\bar q \bar q$ &
$gg \rightarrow q\overline q$ & $q\overline q \rightarrow q\overline q$ &
TOTAL\\ \hline \hline
$\left| y\right| \leq 0.5$ & 1.206 & 0.51 & 0.139 & 0.004 & 1.896 \\ \hline
\end{tabular}
%
\begin{figure}
\vspace{-3cm}
\centerline{\psfig{figure=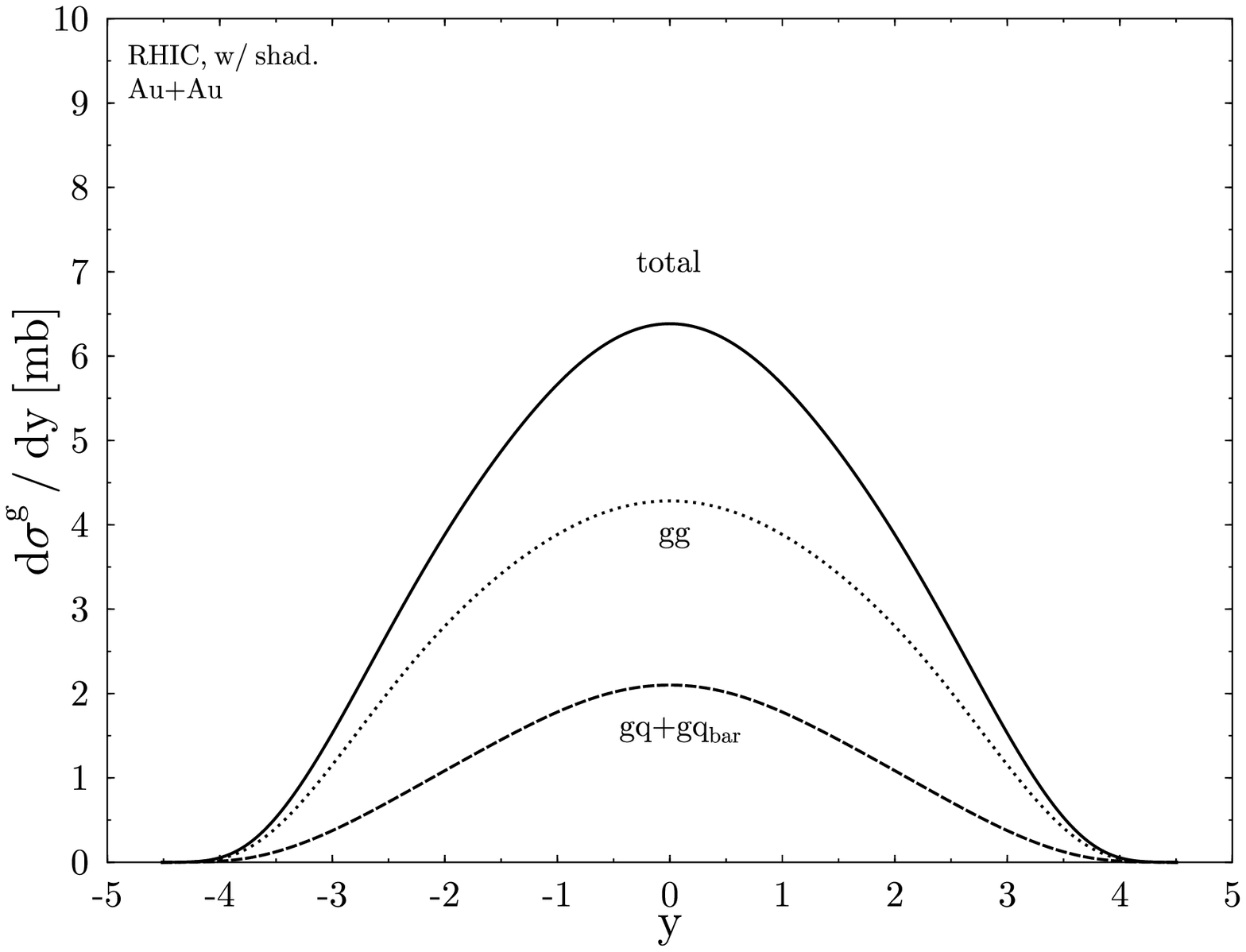,width=11cm}}
\vspace*{-1cm}
\centerline{\psfig{figure=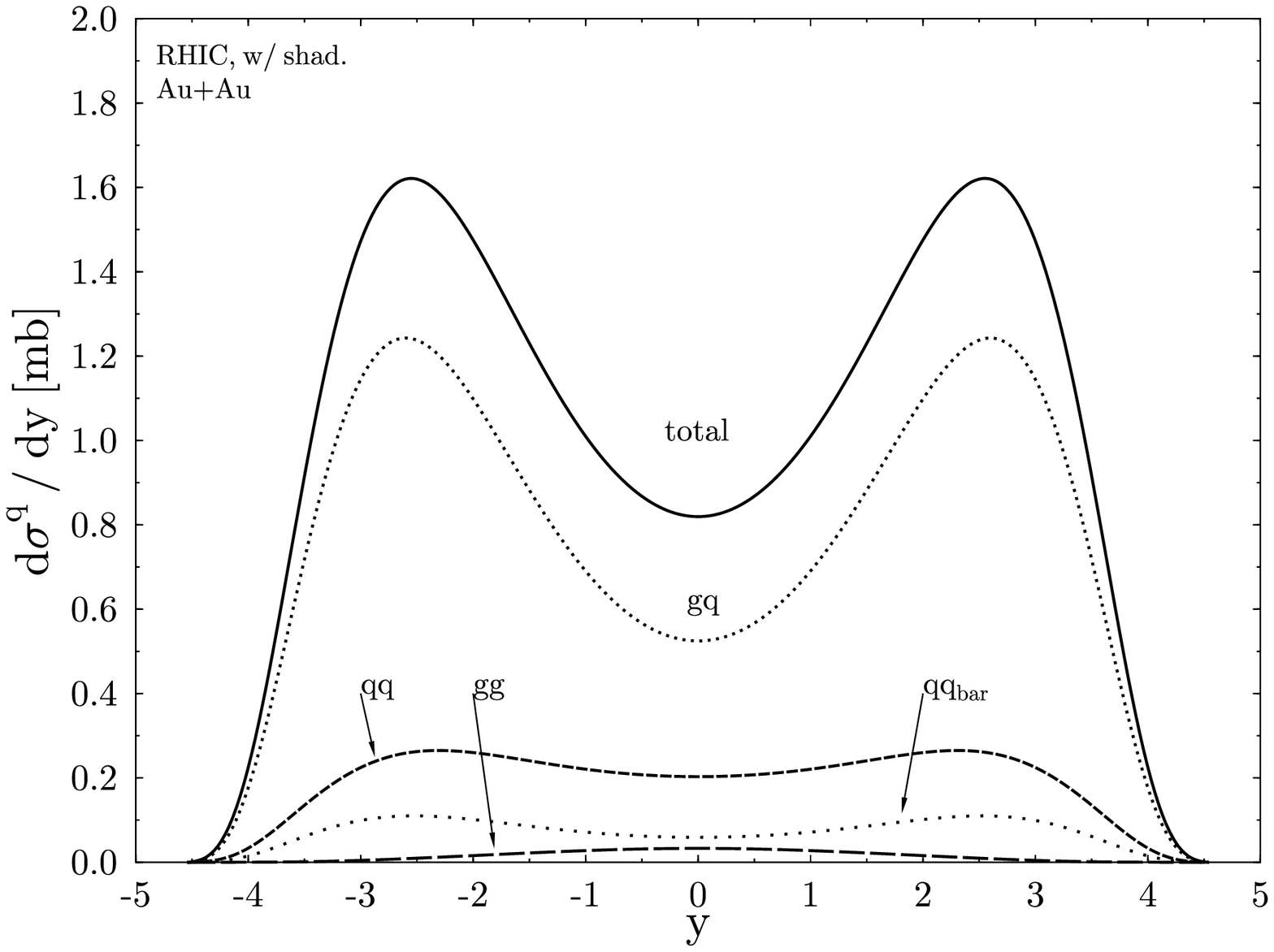,width=11cm}\hspace{-3cm}\psfig{figure=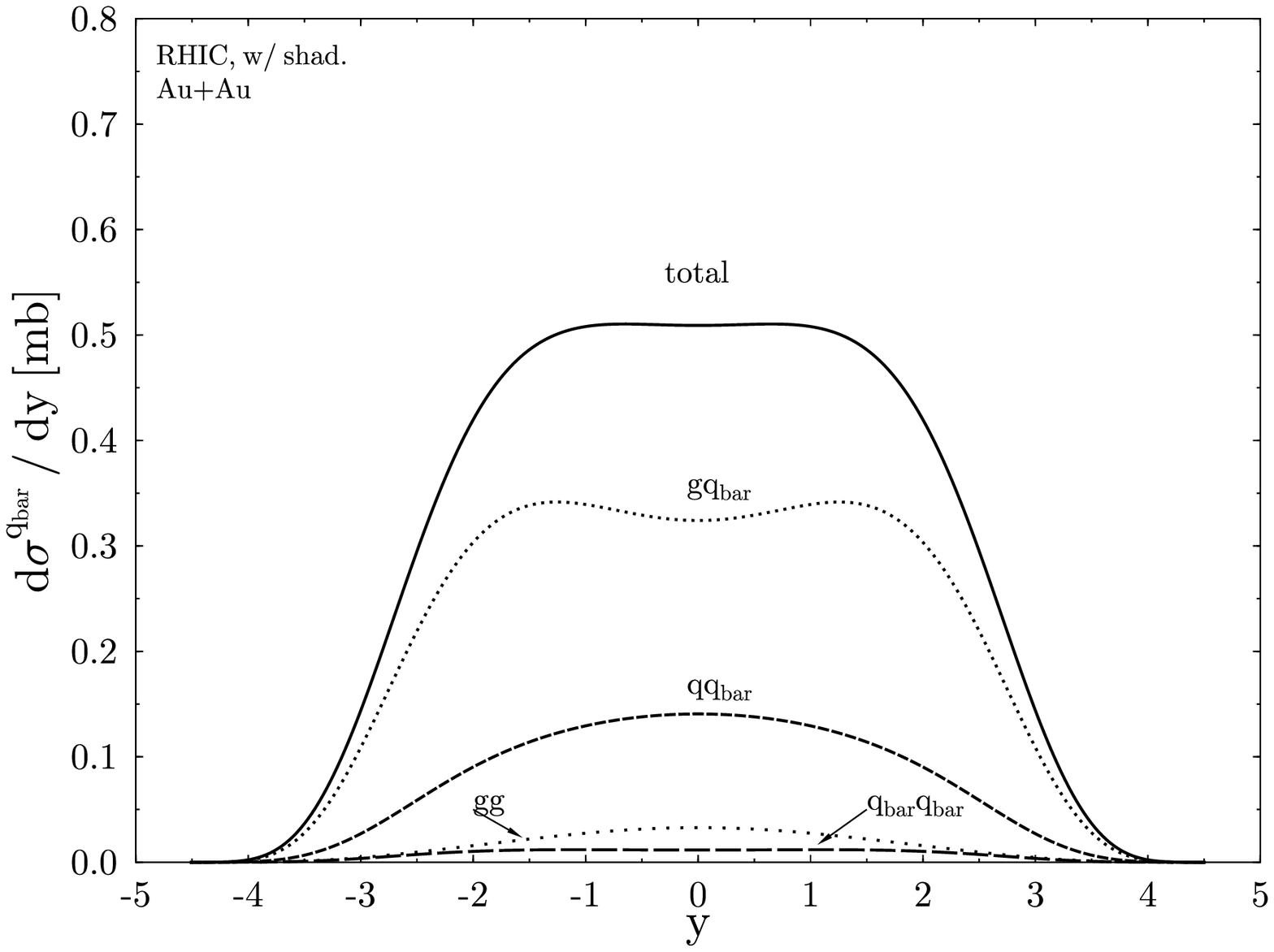,width=11cm}}
\vspace*{-1cm} 
\caption{Rapidity distributions of gluons, quarks, and antiquarks with our modified 
shadowing parametrization shown in figure \protect\ref{nilsshad}.}
\label{fig-rhic-strongshad}
\end{figure}  
\cleardoublepage
For the strong gluon shadowing shown in figure \ref{nilsshad}
one finds the following multiplicities for the different parton species
(the rapidity distributions are shown in figure
\ref{fig-rhic-strongshad}):\\ \\
%
{\bf Table 7:} $\int dy ~dN^g /dy$ {\it for} $\sqrt s =200$ {\it AGeV} \\
\begin{tabular}{|c||c|c|c|} \hline
range of $y$ & $gg\rightarrow gg$ & $gq\rightarrow gq$
+ $g\overline q\rightarrow g\overline q$ & TOTAL\\ \hline \hline
all $y$ & 581.5 & 249.4 & 830.9 \\ \hline
$\left| y\right| \leq 0.5$ & 122.6 & 60.2 & 182.8 \\ \hline
\end{tabular}
\vspace*{1cm}\\
{\bf Table 8:} $\int dy ~dN^q /dy$ {\it for} $\sqrt s =200$ {\it AGeV}\\
\begin{tabular}{|c||c|c|c|c|c|} \hline
range of $y$ & $gq\rightarrow gq$ & $qq\rightarrow qq$ &
$gg \rightarrow q\overline q$ & $q\overline q \rightarrow q\overline q$ &
TOTAL\\ \hline \hline
all $y$ & 196.5 & 48.6 & 3.6 & 18.1 & 266.8 \\ \hline
$\left| y\right| \leq 0.5$ & 15.9 & 5.9 & 0.9 & 1.7 & 24.4 \\ \hline
\end{tabular}
\vspace*{1cm}\\
{\bf Table 9:} $\int dy ~dN^{\bar q} /dy$ {\it for} $\sqrt s =200$ {\it AGeV}\\
\begin{tabular}{|c||c|c|c|c|c|} \hline
range of $y$ & $g\bar q\rightarrow g\bar q$ & $q\bar q\rightarrow q\bar q$ &
$gg \rightarrow q\overline q$ & $\bar q\bar q \rightarrow \bar q\bar q$ &   
TOTAL\\ \hline \hline
all $y$ & 52.9 & 18.1 & 3.6 & 1.8 & 76.4 \\ \hline
$\left| y\right| \leq 0.5$ & 9.4 & 4.1 & 0.9 & 0.3 & 14.7 \\ \hline
\end{tabular}
\vspace*{1cm}\\
The first $E_T$ moments for the reactions including our modified strong gluon 
shadowing are given by:\\
{\bf Table 10:} $\sigma^{g} \left< E_T\right>$ [mb GeV]\\
\begin{tabular}{|c||c|c|c|} \hline
range of $y$ & $gg\rightarrow gg$ & $gq\rightarrow gq$
+ $g\overline q\rightarrow g\overline q$ & TOTAL\\ \hline \hline
$\left| y\right| \leq 0.5$ & 11.87 & 5.93 & 17.8 \\ \hline
\end{tabular}
\vspace*{1cm}\\
\newpage
{\bf Table 11:} $\sigma^{q} \left< E_T\right>$ [mb GeV]\\
\begin{tabular}{|c||c|c|c|c|c|} \hline
range of $y$ & $gq\rightarrow gq$ & $qq\rightarrow qq$ &
$gg \rightarrow q\overline q$ & $q\overline q \rightarrow q\overline q$ &
TOTAL\\ \hline \hline
$\left| y\right| \leq 0.5$ & 0.64 & 0.25 & 0.04 & 0.072 & 1.002 \\ \hline
\end{tabular}
\vspace*{1cm}\\
%
{\bf Table 12:} $\sigma^{\bar q} \left< E_T\right>$ [mb GeV]\\
\begin{tabular}{|c||c|c|c|c|c|} \hline
range of $y$ & $gq\rightarrow gq$ & $qq\rightarrow qq$ &
$gg \rightarrow q\overline q$ & $q\overline q \rightarrow q\overline q$ &
TOTAL\\ \hline \hline
$\left| y\right| \leq 0.5$ & 0.37 & 0.013 & 0.037 & 0.165 & 0.585\\ \hline
\end{tabular}
\vspace*{1cm}\\
We also calculated the multiplicities and first $E_T$ moments by employing the
newest available shadowing parametrization of Eskola et al of ref.~\cite{eskola3}
shown in figure \ref{karisshad}. 
As emphasized
above one should note that the shadowing of gluons in this parametrization is smaller 
than the quark shadowing
since it was tried to stay away from any model dependence and just stick to sum rules
expressing the momentum and baryon number conservation but still assuming that at 
small $x$ ($x\approx 10^{-4}$) the gluon ratio should coincide with the sea quark ratio.
By employing this version
we find the results listet in the following tables and shown in figure 
\ref{fig-rhic-98shad} :\\ \\
{\bf Table 13:} $\int dy~dN^g/dy$ {\it for} $\sqrt s =200$ {\it AGeV}\\
\begin{tabular}{|c||c|c|c|} \hline
range of $y$ & $gg\rightarrow gg$ & $gq\rightarrow gq$
+ $g\overline q\rightarrow g\overline q$ & TOTAL\\ \hline \hline
all $y$ & 969.5 & 350.2 & 1319.7 \\ \hline
$\left| y\right| \leq 0.5$ & 201.8 & 81.9 & 283.7 \\ \hline
\end{tabular}
\vspace*{1cm}\\
{\bf Table 14:} $\int dy ~dN^q/dy$ {\it for} $\sqrt s =200$ {\it AGeV}\\
\begin{tabular}{|c||c|c|c|c|c|} \hline
range of $y$ & $gq\rightarrow gq$ & $qq\rightarrow qq$ &
$gg \rightarrow q\overline q$ & $q\overline q \rightarrow q\overline q$ &
TOTAL\\ \hline \hline
all $y$ &  282.8 & 50.8 & 65.3 & 18.9 & 417.8 \\ \hline
$\left| y\right| \leq 0.5$ & 19.6 & 6.3 & 1.5 & 1.7 & 29.14 \\ \hline
\end{tabular}
\vspace*{1cm}\\
\newpage
{\bf Table 15:} $\int dy ~dN^{\bar q}/dy$ {\it for} $\sqrt s =200$ {\it AGeV}\\
\begin{tabular}{|c||c|c|c|c|c|} \hline
range of $y$ & $g\bar q\rightarrow g\bar q$ & $q\bar q\rightarrow q\bar q$ &
$gg \rightarrow q\overline q$ & $\bar q\bar q \rightarrow \bar q\bar q$ &   
TOTAL\\ \hline \hline
all $y$ & 66.8 & 18.7 & 65.3 & 16.7 & 167.5 \\ \hline
$\left| y\right| \leq 0.5$ & 10.9 & 4.1 & 1.5 & 0.3 & 16.9 \\ \hline
\end{tabular}
%
\begin{figure}
\vspace{-3cm}
\centerline{\psfig{figure=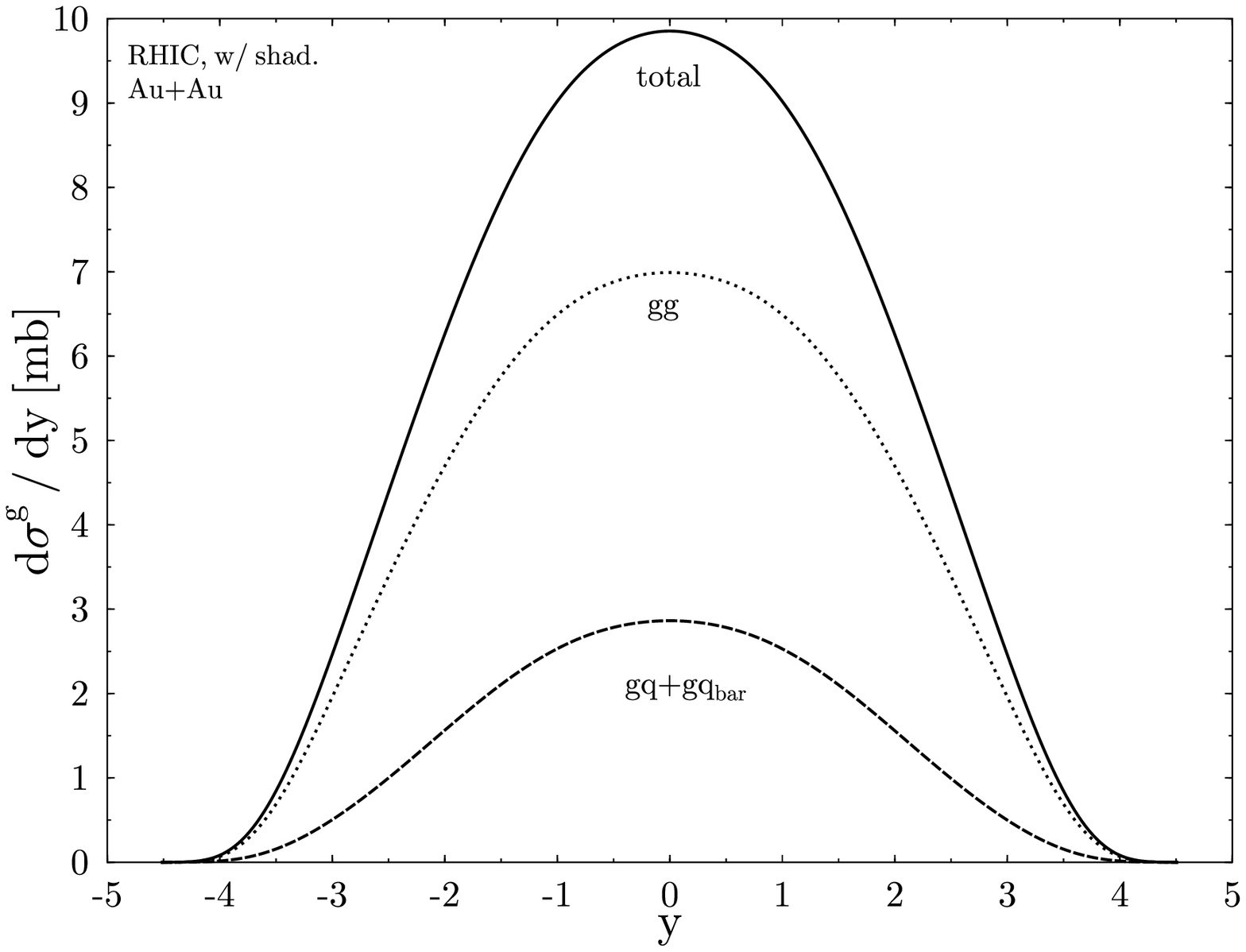,width=11cm}}
\vspace*{-1cm}
\centerline{\psfig{figure=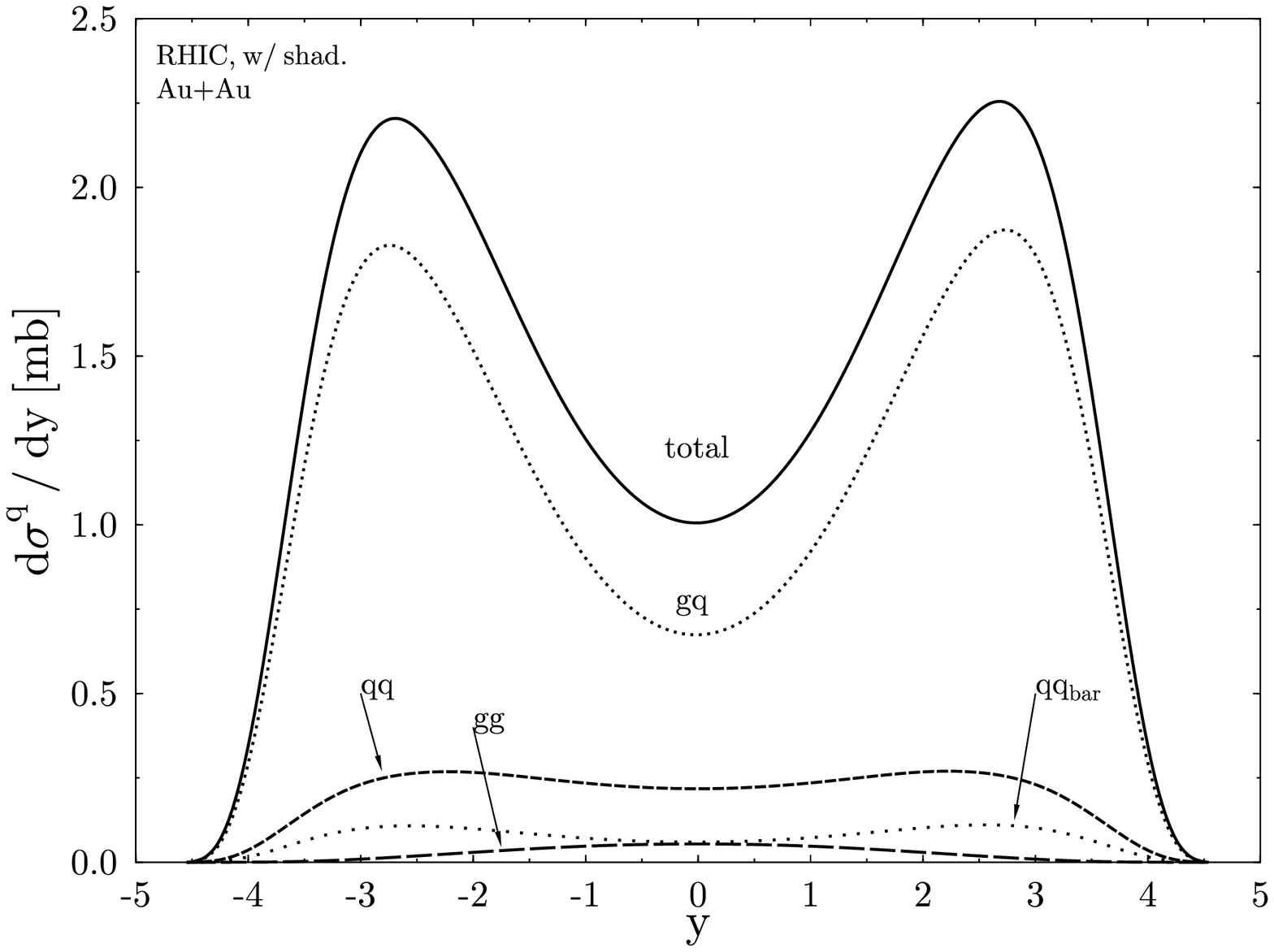,width=11cm}\hspace{-3cm}\psfig{figure=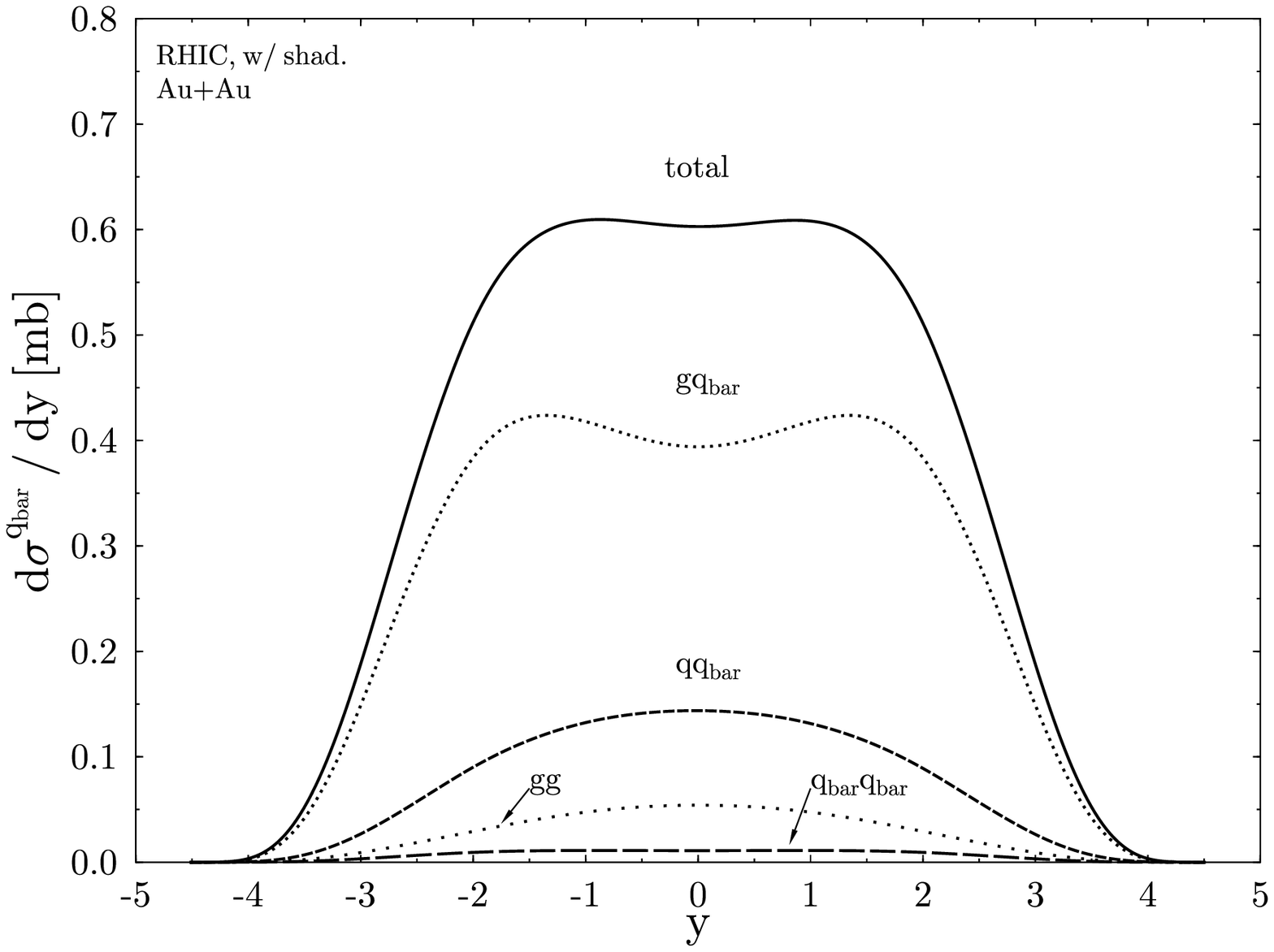,width=11cm}}
\vspace*{-1cm}
\caption{Rapidity distributions of gluons, quarks, and antiquarks with the '98 version of 
the shadowing parametrization shown in figure \protect\ref{karisshad}.}
\label{fig-rhic-98shad}
\end{figure}
\cleardoublepage
In figure \ref{rhic-shadvsnoshad} we directly compared the strong gluon shadowed 
distributions (left figure)
with the unshadowed one. The same was done for the comparison of the $Q^2$ dependent
'98 shadowing version \cite{eskola3} with the unshadowed one (right figure). 
The solid lines give the total contribution, the dotted ones the contribution from the 
$gg$ subprocess and the dashed lines give the $gq+g\bar q$ contribution. The thick 
lines denote the unshadowed distributions and the thin ones the two shadowed ones.
Note that due to the onset of gluon shadowing in the '98 version at such small values 
of $x$ one even gets an enhancement for the $gg\rightarrow gg$ subprocess at RHIC.
We also calculated the $p_T$ distribution without and with the two shadowing versions at 
midrapidity (figure \ref{rhic-ptdist}). Unlike the strong shadowing case the cross over 
point of the curves already happens at $p_T \approx 2.5$ GeV for the '98 gluon shadowing 
version which immediately explains the enhancement in the rapidity distribution.
For the first $E_T$ moment of the transverse energy we find with the shadowing
parametrization of Eskola et al\\
{\bf Table 16:} $\sigma^{g} \left< E_T\right>$ [mb GeV]\\
\begin{tabular}{|c||c|c|c|} \hline
range of $y$ & $gg\rightarrow gg$ & $gq\rightarrow gq$
+ $g\overline q\rightarrow g\overline q$ & TOTAL\\ \hline \hline
$\left| y\right| \leq 0.5$ & 19.2 & 8.05 & 27.25 \\ \hline
\end{tabular}
\vspace*{1cm}\\
\newpage
{\bf Table 17:} $\sigma^{q} \left< E_T\right>$ [mb GeV]\\
\begin{tabular}{|c||c|c|c|c|c|} \hline
range of $y$ & $gq\rightarrow gq$ & $qq\rightarrow qq$ &
$gg \rightarrow q\overline q$ & $q\overline q \rightarrow q\overline q$ &
TOTAL\\ \hline \hline
$\left| y\right| \leq 0.5$ & 2.03 & 0.67 & 0.15 & 0.018 & 2.87 \\ \hline
\end{tabular}
\vspace*{1cm}\\
{\bf Table 18:} $\sigma^{\bar q} \left< E_T\right>$ [mb GeV]\\
\begin{tabular}{|c||c|c|c|c|c|} \hline
range of $y$ & $g\bar q\rightarrow g\bar q$ & $\bar q\bar q\rightarrow \bar q  \bar q$ &
$gg \rightarrow q\overline q$ & $q\overline q \rightarrow q\overline q$ &
TOTAL\\ \hline \hline
$\left| y\right| \leq 0.5$ & 1.123 & 0.032 & 0.148 & 0.423 & 1.726 \\ \hline
\end{tabular}
\vspace*{1cm}\\
From the results above we can calculate the total transverse energy 
$E_T = \sigma <E_T>T_{AA}(0)$ 
carried by the partons,
the number and energy densities $n_f$ and $\varepsilon _f$, and also derive the initial 
temperature $T_i$ if we assume the behavior of an ideal gas of partons. To do so we need
the initial volume. With $R_A = A^{1/3}\cdot 1.1~fm$, $T_{AuAu}(0)=29/mb$, and 
$R_{Au}=6.4~fm$ we find $V_i=\pi R_{A}^{2}\Delta y \tau =12.9~fm^3$.\\
Therefore at RHIC without any shadowing and with K=2.5 we have at midrapidity a total number of
{\bf 284 gluons}, {\bf 32 quarks}, and {\bf 20 antiquarks}. These carry a transverse 
energy of {\bf 774 GeV} (gluons), {\bf 93 GeV} (quarks), and {\bf 55 GeV} (antiquarks).\\
It is then straight forward to derive the number densities by dividing by the initial 
volume to yield: ${\bf n_g = 22~fm^{-3}}$, ${\bf n_q = 2.5~fm^{-3}}$,
${\bf n_{\bar q} = 1.5~fm^{-3}}$. The energy densities can be derived in an analogous 
way to give:\\
${\bf \varepsilon _g = 60~GeV/fm^{-3}}$, 
${\bf \varepsilon _q = 7.2~GeV/fm^{-3}}$, and
${\bf \varepsilon _{\bar q} = 4.3~GeV/fm^{-3}}$. 
If we assume total equilibrium we can derive
the initial temperature from these numbers as 
\begin{equation}
\varepsilon ^{ideal} =  16 \pi ^2 \frac{3}{90}T_{eq}^4
\end{equation}

At this point some comments are appropriate: one could wonder whether the system can be
in equilibrium since one has only hard $2\rightarrow 2$ parton scatterings in this
Glauber approach.  Also one often assumes global equilibrium to be established after, say
1$fm/c$. Now here we are mainly interested in local equilibrium as it is required for
example for hydrodynamical calulcations. The equilibration of partons in a local cell 
happens to be much faster for the following reasons. The high $Q^2$ hard scatterings 
among the partons are absolutely unimportant for the equipartition of longitudinal and transverse
degrees of freedom. It are the soft interactions that are responsible for this feature and
there is a huge resource of soft partons available in the nucleons, even when assuming
the parton distributions to be shadowed in heavy nuclei. The link to the short 
equilibration time is the fact that even though the nucleus is Lorentz contracted to
$L/cosh~y$, the partons obey the uncertainty principle and are therefore smeared out
to distances $1/xP$ in the infinite momentum frame and so the major part of the partons
is {\it outside} the Lorentz contracted disk. Based on some basic priciples and by 
using the Fokker-Planck equation \cite{hwa1,hwa2} the time it takes to establish
local equilibrium in a cell was estimated to have a lower bound of 
$\tau _0 \approx 0.15~fm/c$. As noted above 
we introduced a lower momentum cut-off $p_0 = 2~GeV$ corresponding to a proper time 
of about $0.1~fm/c$. So therefore we may  not be far from local equilibration and the 
calculation on the initial temperature from the initial energy densitiy 
could be rather justified.\\

For the temperature we take into account only the gluons due to their large multiplicity
and energy density that dominates the respective values for the quarks.
We then find ${\bf T_i = 549.52~MeV}$ for RHIC. If we neglect all higher orders, i.e.~take 
a K-factor of K=1 (which of course is wrong, but it is instructive to see the impact on 
$T_i$), we get $T^{K=1}_{i} = 437~MeV$.\\
The same quantities were then calculated for the two different shadowing scenarios.
For the calculations employing the strong gluon shadowing we found that there are
{\bf 183 gluons}, {\bf 25 quarks}, and {\bf 15 antiquarks} carrying transverse energies
of {\bf 516 GeV} (gluons), {\bf 29 GeV} (quarks), and {\bf 17 GeV} (antiquarks).
The resulting number and energy densities are found to be
${\bf n_g = 14~fm^{-3}}$, 
${\bf n_q = 1.9~fm^{-3}}$,
${\bf n_{\bar q} = 1.2~fm^{-3}}$, 
${\bf \varepsilon _g = 40~GeV/fm^{-3}}$, 
${\bf \varepsilon _q = 2.3~GeV/fm^{-3}}$, and 
${\bf \varepsilon _{\bar q} = 1.3~GeV/fm^{-3}}$.
When we calculate the initial temperature for an ideal parton gas from these numbers
we find that the initial temperature decreases due to the reduced number and energy densities 
having their origin in the shadowing of the parton distributions. We find
${\bf T_{i,shad} = 496.5~MeV}$ for a K factor of 2.5 and when neglecting all higher
order contributions we derive $T_{i,shad}^{K=1} = 394.9~MeV$. So what we can learn
here ist the following: due to the reduced number of partons involved in the hard processes
a reduction in the number densities and therefore in the energy densities entering the
formula for the temperature of a thermalized parton gas results. One should note that the
onset of shadowing in our modified shadowing parametrization was chosen same for quarks 
and gluons in accordance with the onset of coherent scattering of a quark antiquark
or gluon gluon pair, respectively off a nucleus. Now in the second shadowing 
parametrization we employed \cite{eskola3} one finds that the onset of shadowing for 
gluons starts at 
smaller momentum fractions from $xG^{Sn}(x)/xG^{C}(x)$ data. With a momentum
cut-off $p_0=2$ GeV the momentum fractions involved in processes at midrapidity are 
bound from below at $x= 0.02$. Therefore one is right on the edge of the onset of 
shadowing of the parametrizations and one should expect the very interesting case that
one is on the edge to the antishadowing region for gluons in the '98 parametrization of 
Eskola et al but not so for the parametrization employing the strong gluon shadowing. 
This behavior is immediately reflected in the number and energy densities.
We found that for this specific shadowing parametrization one has
{\bf 284 gluons}, {\bf 29 quarks}, and {\bf 17 antiquarks} carrying transverse energies
of {\bf 790 GeV} (gluons), {\bf 83 GeV} (quarks), and {\bf 50 GeV} (antiquarks).
We found the following densities:
${\bf n_g = 22~fm^{-3}}$,
${\bf n_q = 2.2~fm^{-3}}$,
${\bf n_{\bar q} = 1.3~fm^{-3}}$,
${\bf \varepsilon _g = 61.2~GeV/fm^{-3}}$,
${\bf \varepsilon _q = 6.43~GeV/fm^{-3}}$, and\\
${\bf \varepsilon _{\bar q} = 3.88~GeV/fm^{-3}}$.
These numbers result in an initial temperature of 
${\bf T_{i,shad} = 552.3~MeV}$ and $T_{i,shad}^{K=1} = 439.2~MeV$, 
respectively.
\\

\begin{figure}
\vspace{-3.5cm}
\centerline{\psfig{figure=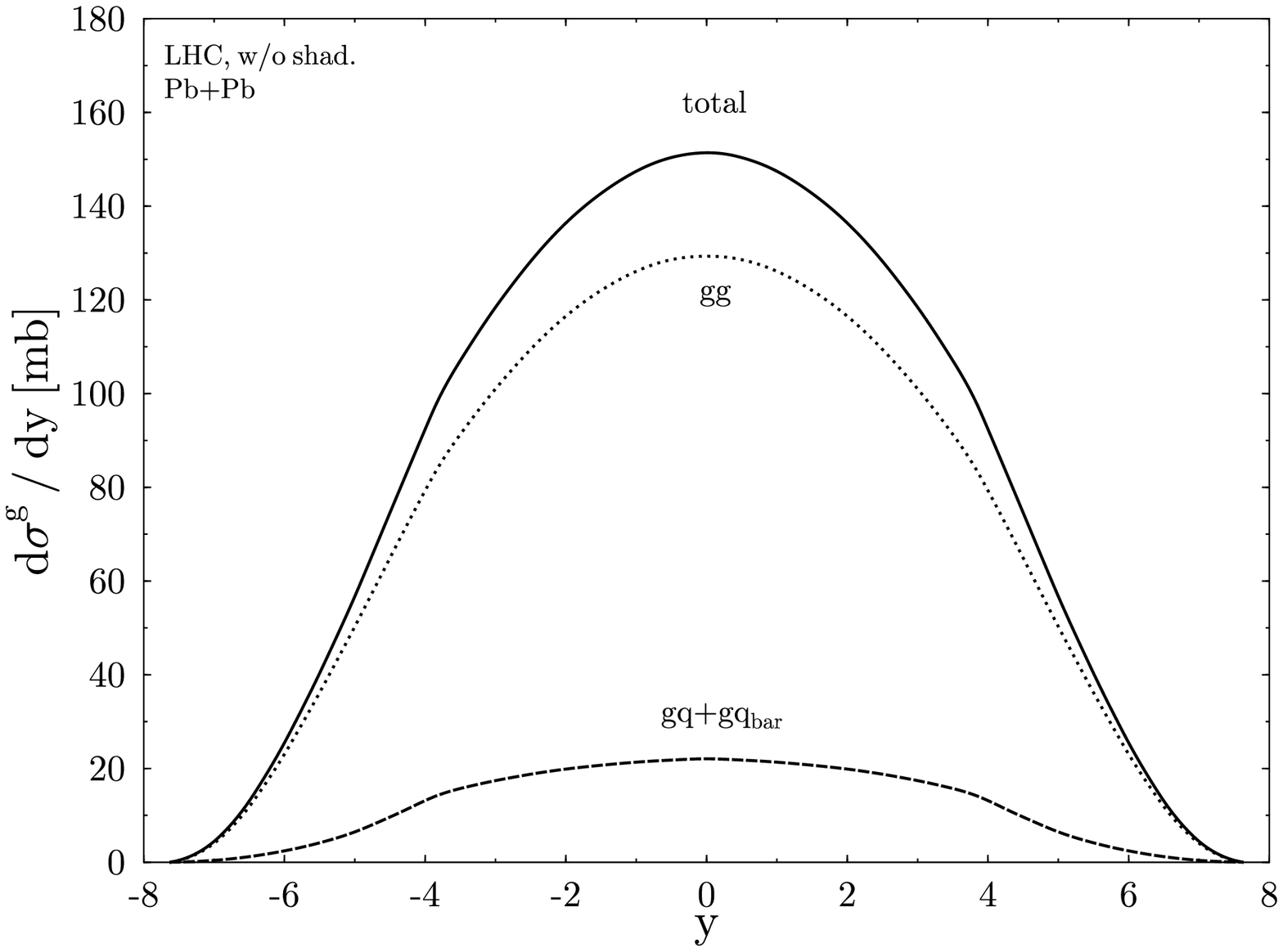,width=11cm}}
\vspace*{-1.5cm}
\centerline{\psfig{figure=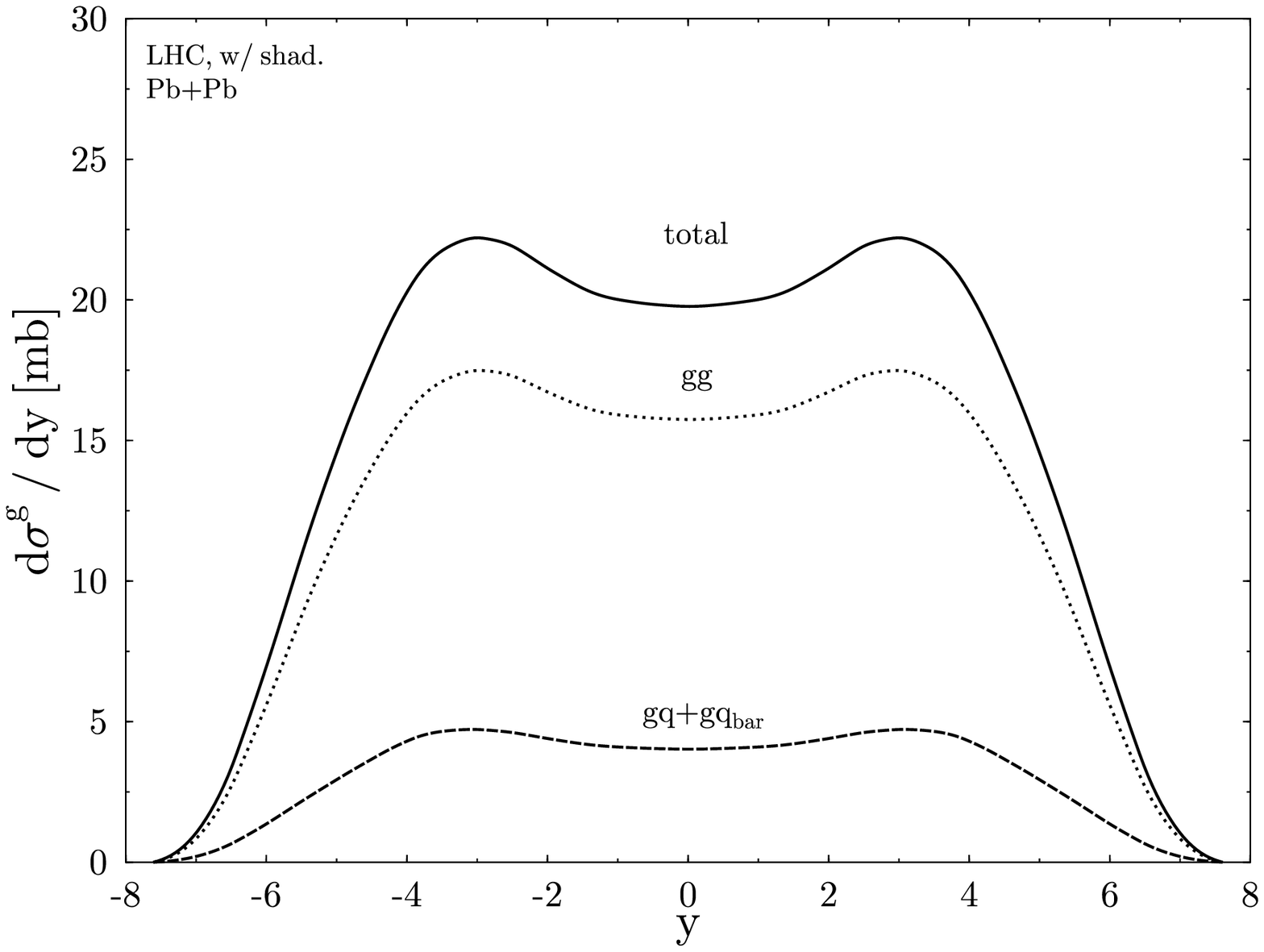,width=11cm}}
\vspace*{-1.5cm}
\centerline{\psfig{figure=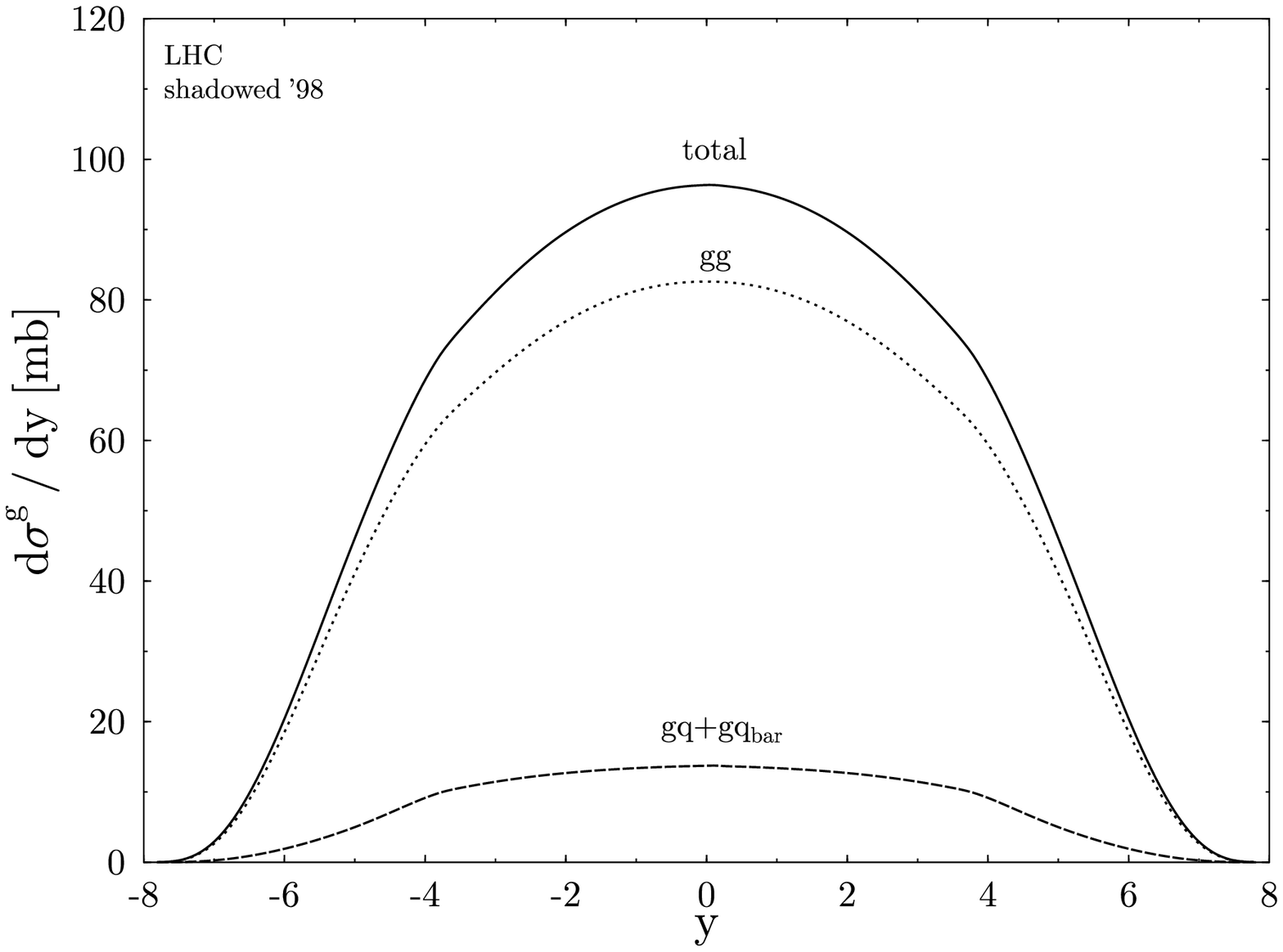,width=11cm}}
\vspace*{-1.5cm}
\caption{Rapidity distributions of unshadowed (upper figure) and shadowed gluons 
(lower two figures) at LHC. The figure in the middle was derived by employing the 
strong gluon shadowing, whereas the bottom figure employed the $Q^2$ dependent
'98 version. Note the change in shape when the strong gluon shadowing is employed.}
\vspace*{-1cm}
\label{fig-lhc-shad}
\end{figure}

\begin{figure}
\vspace{-3cm}
\centerline{\psfig{figure=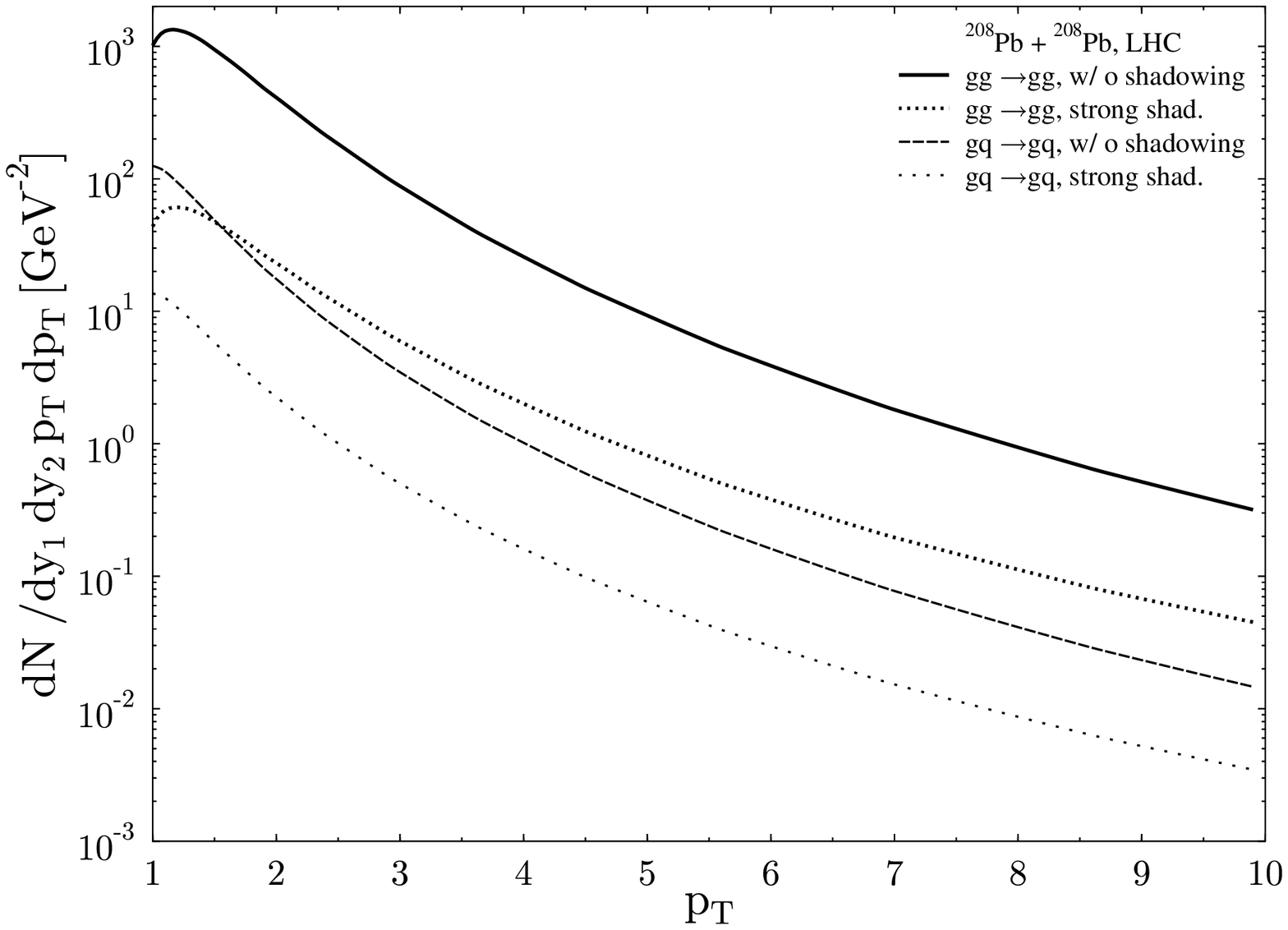,width=11cm}\hspace{-3.2cm}\psfig{figure=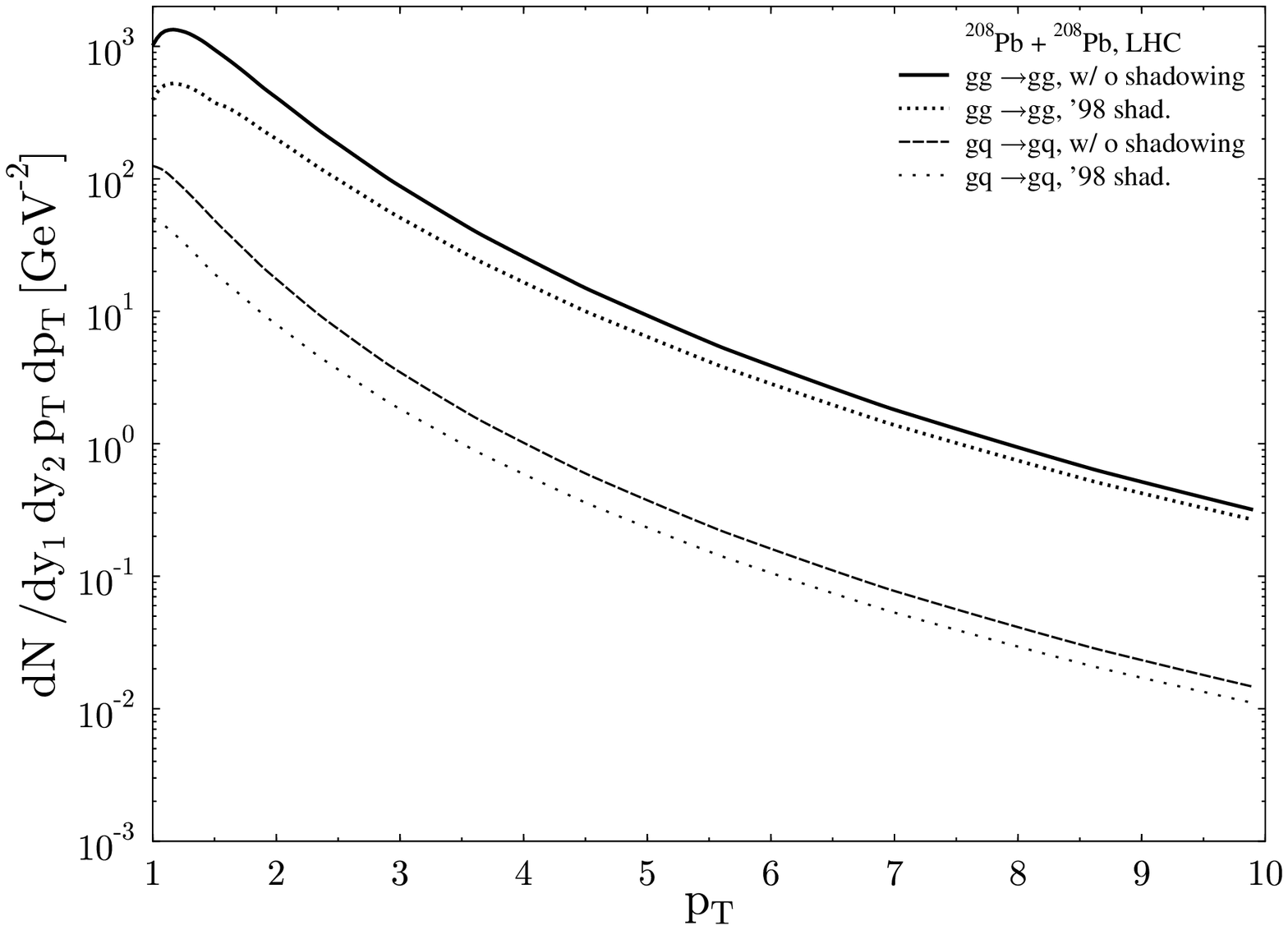,width=11cm}}
\vspace*{-1.5cm}
\caption{$p_T$ distributions for the two shadowing parametrizations at midrapidity 
for LHC.}
\label{lhc-ptdist}
\end{figure}

We also went through the same program to investigate the impact of the different
shadowing parametrizations at the higher LHC energy of $\sqrt s=5.5$ TeV. We here used
the newer parton distributions of CTEQ4L since the involved momentum fractions are so 
small that any new information at small $x$ are valuable. When comparing GRV '94 and CTEQ4L
one finds a difference of about a factor of two at $x\approx 10^{-5}$. At LHC energies the
effect of shadowing should be much more relevant than at RHIC due to the region of smaller
$x$ that gets probed. Because of the strong dominance of the gluon component in the nucleon
we restricted ourself to the calculation of $\sigma ^g$, $\bar N^g$, and therefore on 
the transverse energy and temperature produced by the final state gluons only. Let us first
begin with the unshadowed results.\\ \\
{\bf Table 19:} $\int dy~dN^g/dy$ {\it for} $\sqrt s =5.5$ {\it ATeV}\\
\begin{tabular}{|c||c|c|c|} \hline
range of $y$ & $gg\rightarrow gg$ & $gq\rightarrow gq$
+ $g\overline q\rightarrow g\overline q$ & TOTAL\\ \hline \hline
all $y$ & 36822.7 & 6006.1 & 42828.8\\ \hline
$\left| y\right| \leq 0.5$ & 4137.6 & 707.2 & 4844.8 \\ \hline
\end{tabular}
\vspace*{1cm}\\
%
{\bf Table 20:} $\sigma^{g} \left< E_T\right>$ [mb GeV]\\
\begin{tabular}{|c||c|c|c|} \hline
range of $y$ & $gg\rightarrow gg$ & $gq\rightarrow gq$
+ $g\overline q\rightarrow g\overline q$ & TOTAL\\ \hline \hline
$\left| y\right| \leq 0.5$ & 438.09 & 74.92 & 513.01 \\ \hline
\end{tabular}
\vspace*{1cm}\\
The rapidity distributions for unshadowed and shadowed gluons at LHC is depicted in figure 
\ref{fig-lhc-shad}.
\cleardoublepage
For the strong gluon shadowing we find the following results\\
{\bf Table 21:} $\int dy~dN^g/dy$ {\it for} $\sqrt s =5.5$ {\it ATeV}\\
\begin{tabular}{|c||c|c|c|} \hline
range of $y$ & $gg\rightarrow gg$ & $gq\rightarrow gq$
+ $g\overline q\rightarrow g\overline q$ & TOTAL\\ \hline \hline
all $y$ & 5968.9 & 1558.7 & 7527.6 \\ \hline
$\left| y\right| \leq 0.5$ & 504.9 & 129.3 & 634.2 \\ \hline
\end{tabular}
\vspace*{1cm}\\
{\bf Table 22:} $\sigma^{g} \left< E_T\right>$ [mb GeV]\\
\begin{tabular}{|c||c|c|c|} \hline
range of $y$ & $gg\rightarrow gg$ & $gq\rightarrow gq$
+ $g\overline q\rightarrow g\overline q$ & TOTAL\\ \hline \hline
$\left| y\right| \leq 0.5$ & 48.17 & 12.22 & 60.39 \\ \hline
\end{tabular}\\
\vspace*{1cm}\\
With the weaker gluon shadowing one finds\\
{\bf Table 23:} $\int dy~dN^g/dy$ {\it for} $\sqrt s =5.5$ {\it ATeV}\\
\begin{tabular}{|c||c|c|c|} \hline
range of $y$ & $gg\rightarrow gg$ & $gq\rightarrow gq$
+ $g\overline q\rightarrow g\overline q$ & TOTAL\\ \hline \hline
all $y$ & 24919.1 & 3867.8 & 28786.9 \\ \hline
$\left| y\right| \leq 0.5$ & 2643.2 & 438.4 & 3081.6 \\ \hline
\end{tabular}
\vspace*{1cm}\\
{\bf Table 24:} $\sigma^{g} \left< E_T\right>$ [mb GeV]\\
\begin{tabular}{|c||c|c|c|} \hline
range of $y$ & $gg\rightarrow gg$ & $gq\rightarrow gq$
+ $g\overline q\rightarrow g\overline q$ & TOTAL\\ \hline \hline
$\left| y\right| \leq 0.5$ & 245.92 & 40.95 & 286.87 \\ \hline
\end{tabular}
\vspace*{1cm}\\
\begin{figure}
\vspace{-2cm} 
\centerline{\psfig{figure=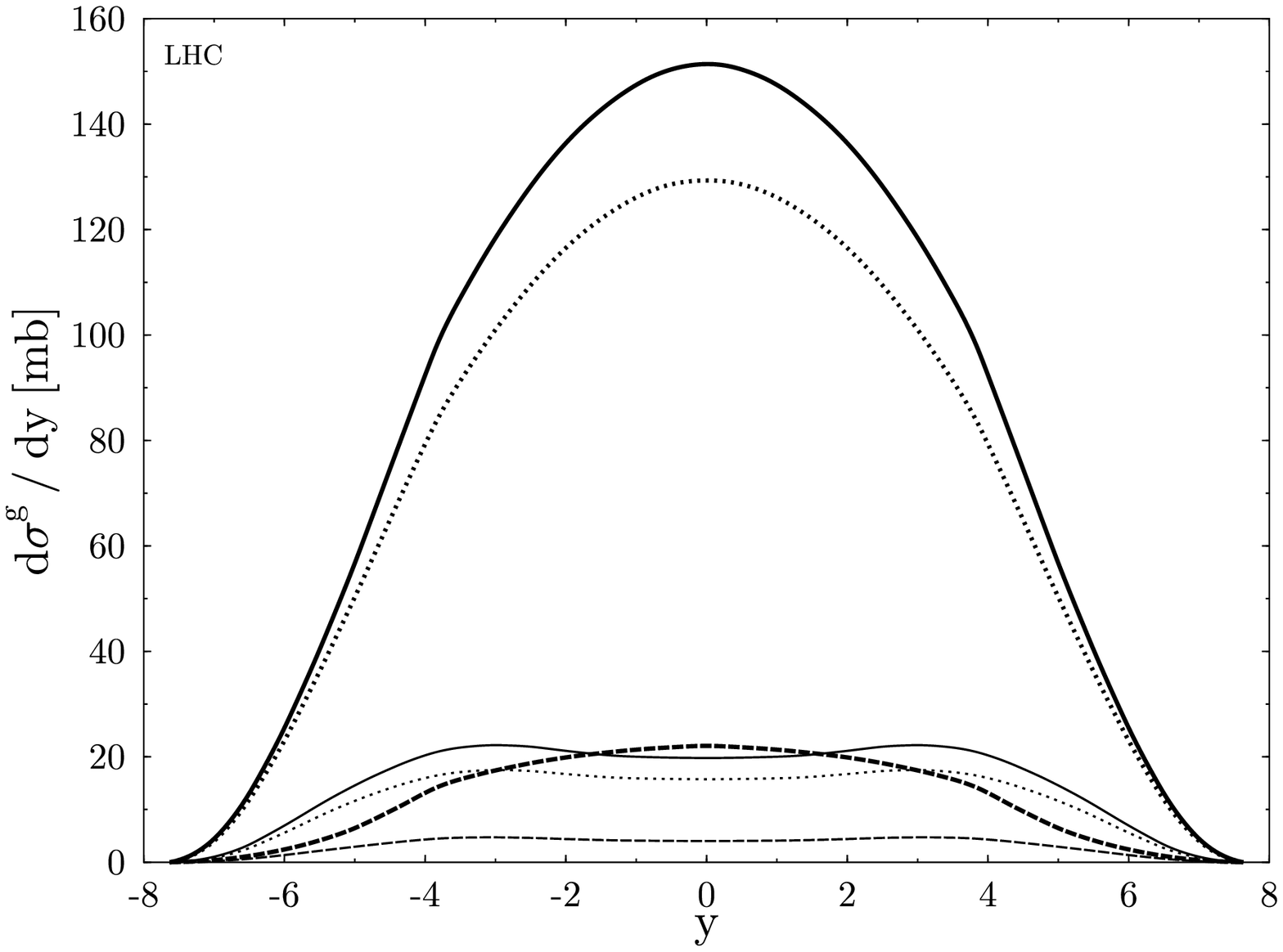,width=11cm}\hspace{-3.2cm}\psfig{figure=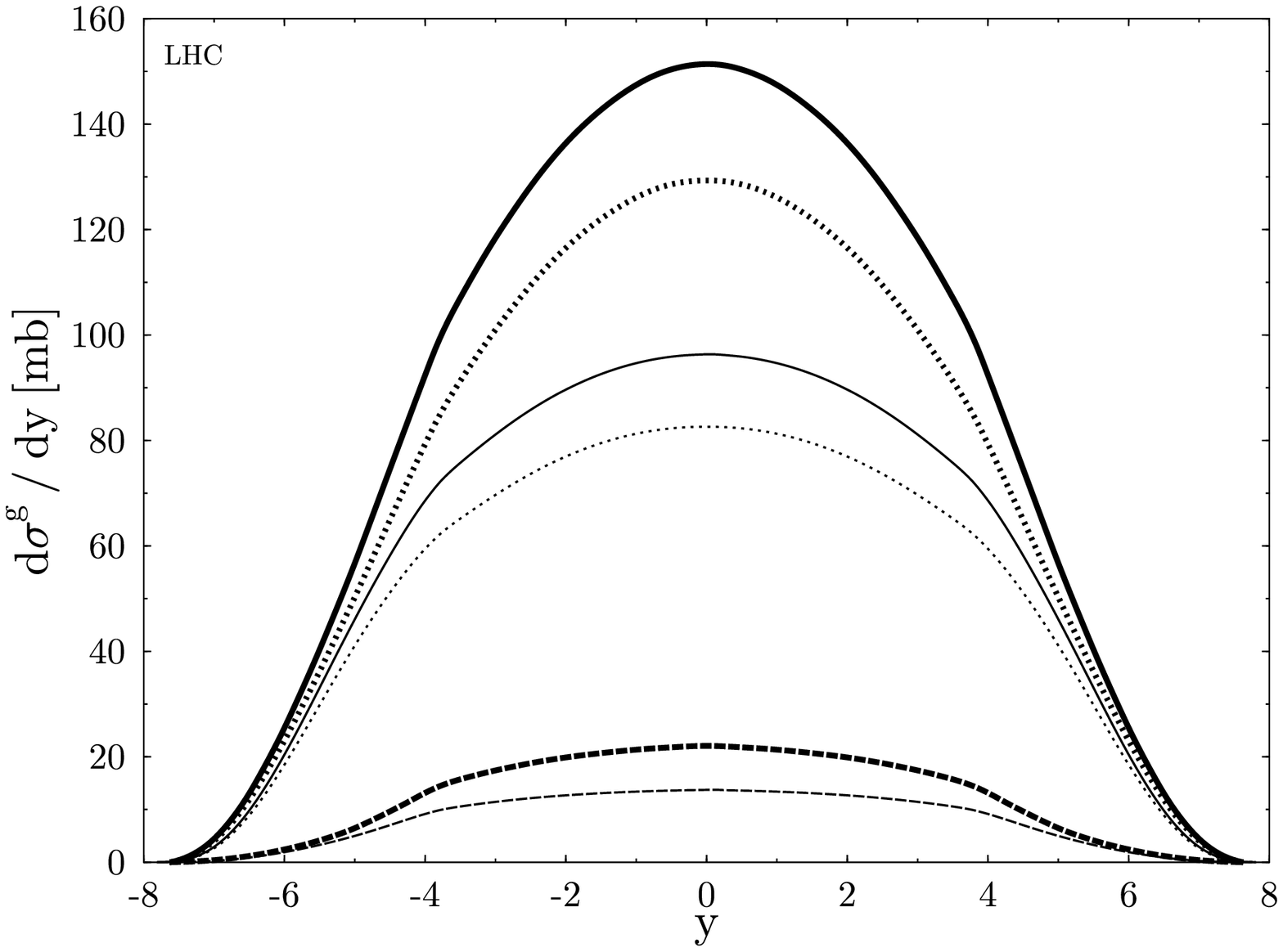,width=11cm}}
\vspace*{-1cm}
\caption{Comparison of the rapidity distributions with strong (left figure)
and weak gluon shadowing (right figure) to unshadowed distribution for LHC. The solid lines
give the total contribution, the dotted ones depict the $gg\rightarrow gg$ process
and the dashed ones stand for the $gq\rightarrow gq$+$g\bar q\rightarrow g\bar q$
processes. The thick lines again give the unshadowed results.}
\label{lhc-shadvsnoshad}
\end{figure}
A direct comparison between the results for shadowed and unshadowed parton distribution
functions is shown in figure \ref{lhc-shadvsnoshad} and
the $p_T$ distributions for LHC are shown in figure \ref{lhc-ptdist}.
\cleardoublepage

Therefore we find the following numbers at LHC: for unshadowed parton distributions one has
at midrapidity {\bf 4845 gluons} that carry a transverse energy of {\bf 16.4 TeV}.
The number density thus is ${\bf n_g = 363~fm^{-3}}$ and the energy density is given by
${\bf \varepsilon _g = 1229.7~GeV/fm^{-3}}$. The initial temperature of an ideal gas derived
with these numbers is ${\bf T_{i} = 1169~MeV}$ and $T_{i}^{K=1.0} = 1056.5~MeV$ 
for K=1.
With the strong gluon shadowing we find {\bf 634 gluons} carrying a transverse energy
of {\bf 1.93 TeV}. We therefore have ${\bf n_g = 47.5~fm^{-3}}$ and
${\bf \varepsilon _g = 144.8~GeV/fm^{-3}}$ resulting in
${\bf T_{i} = 684.9~MeV}$ for K=1.5 and ${\bf T_{i}^{K=1.0} = 618.9~MeV}$. With 
the shadowing version of \cite{eskola3} we find {\bf 3082 gluons} which carry a total 
transverse energy of {\bf 9.18 TeV},
${\bf n_g = 230.9~fm^{-3}}$, and 
${\bf \varepsilon _g = 678.6~GeV/fm^{-3}}$ which results in a temperature
${\bf T_{i} = 1011.08~MeV}$ for K=1.5 and $T_{i}^{K=1.0} = 913.62~MeV$ for K=1.\\
\centerline{\bf 5. Entropy production and $\pi$ multiplicities}
As is known, total entropy and entropy density, respectively, play a very important role 
in the formation of a quark-gluon plasma. Total entropy reaches its final value when 
the system equilibrates and can, if assuming
an adiabatical further evolution, be related to the effective number of degrees of 
freedom in the quark-gluon and in a pure pion plasma via \cite{bjorken,kkg}
\begin{equation}
r=\frac{s^\pi (T_c)}{s^{qg} (T_c)}\approx 0.7\pm 0.2
\end{equation}
where $s^\pi$ and $s^{qg}$ are the entropy densities in the pion and quark-gluon plasma.
The total entropy can then be related to the pion multiplicity as 
\begin{equation}
\frac{dS}{dy}=c^{qg} \left( \frac{dN^{qg}}{dy} \right)_{b=0}
\approx \frac{c^\pi}{r}\left( \frac{dN^\pi}{dy} \right)_{b=0}
\end{equation}
where $c^{qg}=4.02$ for $N_f =4$ and $c^\pi \approx 3.6$.\\
A note on the separation between hard and soft processes is appropriate at this point.
As emphasized above we introduced a cut-off at $p_0 = 2$ GeV to ensure the applicability
of perturbative QCD. Nevertheless there is always a soft component contributing to the
production of transverse energy neglected in our studies so far. In \cite{eskola2}
it was shown that with $p_0 = 2$ GeV at SPS the hard partons only carry about
$4\%$ of the total transverse energy $E_T$. At RHIC energies they carry $\approx 50\%$ and
for $\sqrt s = 2$ TeV the hard partons already carry $\approx 80\%$ of the total transverse 
energy. Since we here solely want to investigate the role of shadowing in hard reactions
we will not calculate the pion multiplicity for RHIC where the soft contribution still is
significant but restrict ourselves to the pion number at $y\approx 0$ for LHC energies.\\
If we employ the numbers for the entropy densities in the different plasmas and use 
our findings on the contributions of shadowing to the number of minijets we find that
at $y=0$ one has
\begin{eqnarray}
\left( \frac{dN^\pi}{dy} \right)_{b=0} \approx 3786,\nonumber\\
\left( \frac{dN^\pi}{dy} \right)_{b=0} \approx 2413,\\
\left( \frac{dN^\pi}{dy} \right)_{b=0} \approx 330,\nonumber
\end{eqnarray}
when employing no shadowing, the '98 version of Eskola et al, and the strong gluon shadowing
parametrization.\\
\centerline{\bf 6. Conclusions}
In this paper we investigated the influence of nuclear shadowing on rapidity spectra,
transverse energy production and on macroscopic quantities such as the initial temperature.
We employed two different versions of parametrizations for the shadowing: one with
a strong initial gluon shadowing and a model independent one recently published by
Eskola et al \cite{eskola3}. We found that the latter one gives an enhancement of minijet
production at RHIC in contrast to the other case were a reduction to $\approx 65\%$
results. This difference directly manifests itself in the initial temperature $T_i$ which
happens to be smaller only for the strong gluon shadowing. At LHC the situation changes since
there also the weakly shadowed gluons finally result in lower spectra and $T_i$. Since
the two shadowing parametrizations differ so drastically one finds a large difference
in the results for the number of minijets at midrapidity: for the strong shadowing one 
has $\approx 630$ gluons whereas for the weaker shadowing one finds $\approx 3000$ 
gluons. Since there are so few gluons for the strong gluon shadowing we find that 
the initial temperature at LHC is not dramatically higher than at RHIC!
\centerline{\bf Acknowledgements}
We would like to thank K.J. Eskola for many stimulating discussions on 
minijet production and nuclear shadowing.


\vskip 1.0 cm
\begin{thebibliography}{99}
\bibitem{hwa1} R. Hwa and K. Kajantie, {\it Phys. Rev. Lett.} {\bf 56} (1986) 696
\bibitem{ua1} UA1 Collaboration, C. Albaja et al., {\it Nucl. Phys.} {\bf B309} (1988) 405
\bibitem{eskola1} K.J. Eskola, K. Kajantie, Z. Phys. {\bf C 75} (1997) 515
\bibitem{nils1} N. Hammon, H. St\"ocker, W. Greiner, {\it hep-ph/9811242}, 
accepted for publication in {\it Phys. Lett.} {\bf B} 
\bibitem{grv} M. Gl\"uck, E. Reya, A. Vogt, {\it Z. Phys.} {\bf C67} (1995) 433 
\bibitem{cteq} H.L. Lai, J. Huston, S. Kuhlmann, F. Olness, J. Owens, D. Soper, W.K. Tung, 
H. Weerts, {\it Phys. Rev.} {\bf D55} (1997) 1280
\bibitem{eskola2} K.J. Eskola, K. Kajantie, J. Lindfors, {\it Nucl. Phys.} {\bf B323} (1989) 37
\bibitem{bauer} T. Bauer, R. Spital, D. Yennie, F. Pipkin, {\it Rev. Mod. Phys.}
{\bf 50} (1978) 261
\bibitem{lonya1} B. Blattel, G. Baym, L. Frankfurt, M. Strikman, {\it Phys. Rev. Lett.}
{\bf 70} (1993) 896\\
L. Frankfurt, G. Miller, M. Strikman, {\it Phys. Lett.} {\bf B304} (1993) 1
\bibitem{gribov} V. N. Gribov, {\it JETP} {\bf 30} (1970) 709,\\
R. Glauber in {\it Lectures in theoretical physics}, ed. W.E. Brittin {\it et al.} (1959)
\bibitem{close} F.E. Close, J. Qiu, R.G. Roberts, {\it Phys. Rev.} {\bf D40} (1989) 2820
\bibitem{glr} L. Gribov, E. Levin, M. Ryskin, {\it Phys. Rep.} {\bf 100} (1983) 1
\bibitem{mq} A.H. Mueller, J. Qiu, {\it Nucl. Phys.} {\bf B268} (1986) 427
\bibitem{eskola5} K.J. Eskola, J. Qiu, X. Wang, {\it Phys. Rev. Lett.} {\bf 72} (1994) 36
\bibitem{eskola3} K.J. Eskola, V.J. Kolhinen and C.A. Salgado, JYFL-8/98, US-FT/14-98, 
{\it hep-ph/9807297}\\
K.J. Eskola, V.J. Kolhinen and P.V. Ruuskanen,
"Scale evolution of nuclear parton distributions"
CERN-TH/97-345, JYFL-2/98, hep-ph/9802350
\bibitem{eskola4} K.J. Eskola, {\it Nucl. Phys.} {\bf B400} (1993) 240
\bibitem{gousett} T. Gousett, H.J. Pirner, {\it Phys. Lett.} {\bf B375} (1996) 349
\bibitem{nmc} A. M\"ucklich, Ph.D. thesis, University of Heidelberg, 1995
\bibitem{hwa2} R. Hwa, {\it Phys. Rev.} {\bf D32} (1985) 637
\bibitem{bjorken} J.D. Bjorken, {\it Phys. Rev.} {\bf D27} (1983) 140
\bibitem{kkg} K. Geiger, {\it Phys. Rev.} {\bf D46} (1992) 4986
\end{thebibliography}
\end{document}